\documentclass[aps,prx,floatfix,twocolumn,superscriptaddress,longbibliography]{revtex4-1}
\usepackage{amsmath,amsfonts,amssymb}
\usepackage{graphicx}
\usepackage{epsfig}
\usepackage{dcolumn}
\usepackage{bm}
\usepackage{times}
\usepackage{color}
\usepackage{paralist}

\def\<{\langle}
\def\>{\rangle}
\def\(({\left(}
\def\)){\right)}
\def\[[{\left[}
\def\]]{\right]}

\newcommand{\be}{\begin{equation}}
\newcommand{\ee}{\end{equation}}
\newcommand{\bea}{\begin{eqnarray}}
\newcommand{\eea}{\end{eqnarray}}

\newcommand{\cIT}{c_\text{\tiny IT}}
\newcommand{\cAlg}{c_\text{alg}}
\newcommand{\cBP}{c_\text{\tiny BP}}
\newcommand{\cSA}{c_\text{\tiny SA}}
\newcommand{\cRSA}{c_\text{\tiny RSA}}
\newcommand{\cKS}{c_\text{\tiny KS}}
\newcommand{\TKS}{T_\text{\tiny KS}}
\newcommand{\TB}{T_\text{\tiny Bayes}}
\newcommand{\Tjump}{T_\text{jump}}
\newcommand{\bs}{{\bm s}}

\begin{document}

\title{Limits and performances of algorithms based on simulated annealing\\
in solving sparse hard inference problems}

\author{Maria Chiara Angelini}

\affiliation{Dipartimento di Fisica, Sapienza Universit\`a di Roma, 00185 Italy}
\affiliation{INFN - Sezione di Roma 1, 00185 Italy}

\author{Federico Ricci-Tersenghi}

\affiliation{Dipartimento di Fisica, Sapienza Universit\`a di Roma, 00185 Italy}
\affiliation{INFN - Sezione di Roma 1, 00185 Italy}
\affiliation{CNR - Nanotec, unit\`a di Roma, 00185 Italy}

\begin{abstract}
The planted coloring problem is a prototypical inference problem for which thresholds for Bayes optimal algorithms, like Belief Propagation (BP), can be computed analytically. In this paper, we analyze the limits and performances of the Simulated Annealing (SA), a Monte Carlo-based algorithm that is more general and robust than BP, and thus of broader applicability.
We show that SA is sub-optimal in the recovery of the planted solution because it gets attracted by glassy states that, instead, do not influence the BP algorithm.
At variance with previous conjectures, we propose an analytic estimation for the SA algorithmic threshold by comparing the spinodal point of the paramagnetic phase and the dynamical critical temperature.
This is a fundamental connection between thermodynamical phase transitions and out of equilibrium behavior of Glauber dynamics.
We also study an improved version of SA, called replicated SA (RSA), where several weakly coupled replicas are cooled down together.
We show numerical evidence that the algorithmic threshold for the RSA coincides with the Bayes optimal one.
Finally, we develop an approximated analytical theory explaining the optimal performances of RSA and predicting the location of the transition towards the planted solution in the limit of a very large number of replicas.
Our results for RSA support the idea that mismatching the parameters in the prior with respect to those of the generative model may produce an algorithm that is optimal and very robust. 
\end{abstract}


\maketitle

\section{Introduction}

Inference problems are widespread in all scientific disciplines and real-world applications based on data analysis.
It is also well known that not all inference problems have the same difficulty, and the solving algorithms need possibly to be tested on the hardest problems.
For this reason, focusing on classes of inference problems that are provably very hard is mandatory at the time of testing algorithms.
Here we choose to work with inference problems defined on sparse random graphs that present the so-called \emph{hard phase}, that is a range of the signal-to-noise ratio (SNR) where the solution is in principle achievable (by unbounded computational power), but any known algorithm running in a time growing polynomially with the problem size is not able to reach such a solution. Below we define more in detail this class of hard inference problems; for the moment it is enough to stress that the presence of the hard phase is related to the so-called \emph{information to computational gap conjecture}. This conjecture, partially proved for some inference problems and some classes of algorithms, assumes that the minimal SNR needed by any polynomial-time algorithm to solve one of these hard inference problems is larger than the information theoretical lower bound \cite{holland1983stochastic,decelle2011inference,richard2014statistical,abbe2015detection,deshpande2015finding,lesieur2015mmse,chen2016statistical,hopkins2016fast,bandeira2018notes,barak2019nearly}.

\begin{figure}[h]
    \centering
    \includegraphics[width=\columnwidth]{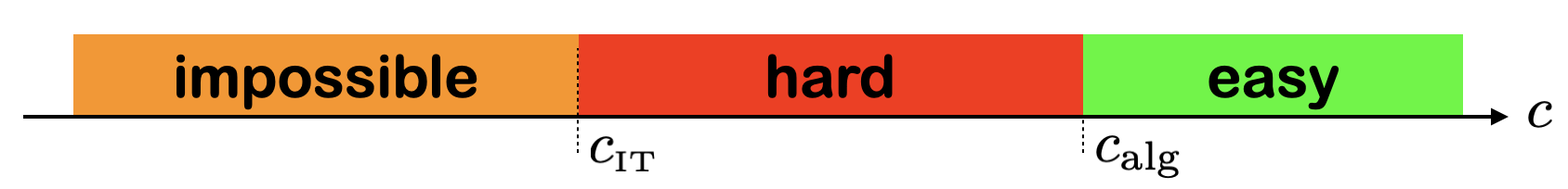}
    \caption{The schematic phase diagram for hard inference problems showing an information-to-computational gap. The SNR is here given by the graph mean degree $c$. While for $c<\cIT$ it is impossible to detect the planted signal, for $c>\cIT$ it is in principle possible, but any known algorithm requires a much larger SNR, $c>\cAlg$, to achieve signal detection. The value of $\cAlg$ depends on the algorithm and the present work aims at estimating it for simulated annealing algorithms.}
    \label{fig:plot0}
\end{figure}

In Fig.~\ref{fig:plot0} we show a schematic phase diagram for a generic inference problem with a hard phase.
The derivation of such a phase diagram has been possible working in the very commonly used setting of Bayesian inference known as the \emph{teacher-student scenario}, firstly introduced in Ref.~\cite{gardner1989three}.
In this setting the teacher generates a random signal $\mathbf{x}$ with the prior probability distribution $P_p(\mathbf{x})$. The signal generated by the teacher is often called the \emph{planted} configuration. Later, the teacher generates the data from the signal using a probabilistic model $P_m$.
We assume that the data come into the form of a sparse graph $G$ and that the SNR is directly related to the graph mean connectivity $c$
\footnote{There are problems in which the data are not just the graph, and the SNR is more complicated than just $c$. However the following reasoning based on Bayes theorem can be generalized to other situations \cite{zdeborova2016statistical}.}.
Applying a probabilistic model to a random signal, we get a random graph $G$ generated with the likelihood $P_m(G|\mathbf{x})$. The teacher provides the student with the prior $P_p$, the model $P_m$ and the data $G$, asking the student to infer the planted signal $\mathbf{x}$. Using the Bayes theorem, the student can easily write down the posterior probability distribution
\begin{equation}
 P(\mathbf{x}|G) \propto P_m(G|\mathbf{x})P_p(\mathbf{x})\;.
\end{equation}
The Bayes optimal estimator $\hat{\mathbf{x}}_\text{opt}$ of the signal can be obtained by sampling the posterior $P(\mathbf{x}|G)$. A weak recovery of the signal can be achieved if the Bayes optimal estimator $\hat{\mathbf{x}}_\text{opt}$ is better than a random guess: this happens for $c>\cIT$, that is above the information theoretical threshold $\cIT$.

In hard inference problems, it is not enough to have an SNR larger than the $\cIT$ threshold to recover the signal in a time growing polynomially with the problem size. A larger algorithmic threshold exists, $\cAlg$, such that only for $c>\cAlg$ polynomial-time algorithms can actually detect the signal. The threshold $\cAlg$ clearly depends on the algorithm, and we are going to show how to estimate it for stochastic samplers based on the Simulated Annealing (SA) algorithm.

When the graph $G$ representing the data is a random graph \cite{erdos1959random}, the posterior can be efficiently sampled by using the Bethe approximation and the Belief Propagation (BP) algorithm \cite{mezard2009information}. The algorithmic threshold for the BP algorithm $\cBP$ is larger than $\cIT$ for inference problems with a hard phase, and is nonetheless considered the optimal algorithmic threshold for algorithms running in linear time in the problem size $N$ in the vast majority of the problems \cite{zdeborova2016statistical} \footnote{There are some counterexamples like planted XORSAT
\cite{franz2001ferromagnet, zdeborova2011quiet} or noiseless sign retrieval with Gaussian matrices \cite{gamarnik2021inference, song2021cryptographic} in which there exist algorithms that do beat BP (even though these examples are very brittle and break down when adding a very small noise)}.
Other algorithms which are not Bayes optimal, i.e.\ not sampling exactly the Bayesian posterior, may have even larger algorithmic thresholds.

Unfortunately, the BP algorithm is known to be not very robust and of limited applicability when the graph $G$ is not locally tree-like, as the presence of short loops may avoid convergence to a fixed point and, even at convergence, the estimated marginal probabilities may be far from the exact ones \cite{yedidia2003understanding}.
We are not saying BP can never be used for inference away from random graphs: indeed many works are showing the applicability of BP or inference methods strongly related to it on real-world data (see \cite{decelle2011asymptotic} just as an example).
However, having at hand more robust algorithms that can work efficiently both on random graphs and on more general structured data is of primary importance.

To this aim, it is natural to consider samplers of the posterior probability distribution based on Monte Carlo Markov Chain (MCMC). By construction, an MCMC will eventually sample exactly from the posterior, but its limits and performances are mostly unknown. 
In some recent works, it has been proved how the algorithmic threshold of some MCMC algorithms is higher than the algorithmic thresholds of other Bayes-optimal algorithms in specific planted problems \cite{gamarnik2019landscape, angelini2021mismatching, chen2022almost}. 
However, apart from the straightforward statement that MCMC methods are slowed down in a phase where ergodicity is broken in the large $N$ limit \cite{antenucci2019glassy}, more precise general statements on when and why MCMC-based inference algorithms can find the solution are missing to the best of our knowledge.

MCMC-based inference algorithms could really be ideal if, on top of being robust, they would also be optimal in those cases where the optimal threshold can be analytically computed. To understand this last point, we believe that the study of inference problems defined on random sparse graphs is the perfect framework: thresholds can be computed analytically thanks to the Bethe approximation for the posterior distribution and MCMC simulations can be run efficiently on very large problem sizes since
a single MC step is linear in $N$ for sparse models.

We study one of the simplest and most common inference problems, namely the planted coloring problem (to be defined in more detail below), using simple, but effective, MCMC-based algorithms: Simulated Annealing (SA) and Replicated Simulated Annealing (RSA).
The main results that we achieve within this framework are the following.
\begin{description}
\item[Algorithmic threshold for Simulated Annealing]\hfill\\
We make a conjecture based on firm numerical evidence and strong analytical arguments for the determination of the algorithmic threshold of SA.
Our conjecture is based on the comparison of the spinodal point and the dynamical phase transition.
Sub-optimal with respect to the Bayes-optimal threshold, we find that the SA algorithmic threshold is worse than what has been reported in previous studies.
\item[Replicated Simulated Annealing is Bayes-optimal]\hfill\\
We provide strong numerical evidence that the algorithmic threshold for RSA is extremely close to the Bayes-optimal one of BP. RSA works very well even for a small number of coupled replicas, so the complexity overhead with respect to SA is minimal.
\item[Nature of the phase transition]\hfill\\
We provide both numerical and analytical evidence that coupling replicas changes the nature of the glass transition in the random problem, from a discontinuous one to a continuous one. This observation may have important algorithmic consequences, as hard phases are known to exist only in models undergoing discontinuous phase transitions.
\item[Analytical estimate for the RSA critical temperature]\hfill\\
We provide an analytical estimate of the transition temperature in the limit of a large number of coupled replicas for the RSA algorithm. Convergence to this limiting value is fast, so it provides a good estimate of the critical temperature also for a reasonably small number of replicas, those used in practical simulations.
\item[Optimal temperature for MCMC is not Bayes optimal]\hfill\\
For the sake of simplicity, we study here an inference problem without additional noise, i.e.\ generated from a Gibbs distribution at $T=0$. We provide analytical evidence that the optimal temperature for recovering the signal is strictly positive and not small at all, so far from the Bayes optimal one (this is in agreement with previous numerical results).
\item[Analytical control of finite size effects]\hfill\\
Analytical predictions in the large $N$ limit are of scarce practical utility if they are not complemented with the study of finite size effect, which may be very large. 
We provide an analytical explanation of finite size effects in the behavior of the SA algorithm, via the study of BP on finite graphs, where similar finite size effects appear.
\end{description}

Taken all together the above results provide a strong and coherent theory explaining the behavior of important algorithms based on Simulated Annealing in solving hard inference problems defined on sparse graphs. We believe this theory has broad applicability, and preliminary results in other planted models support it (however we leave the study of other problems for future works, the planted coloring being already very interesting and of broad applicability \textit{per se}).

The rest of the paper is organized as follows. In Sec.~\ref{sec:plantedCol} we define the planted coloring problem and the associated BP algorithm. In Sec.~\ref{sec:SA} we study the performances of SA and put forward a conjecture for its algorithmic threshold. In Sec.~\ref{sec:RSA} we describe the RSA algorithm and we show that it can reach the threshold of Bayes-optimal algorithms. In Sec.~\ref{sec:FSE} we describe the finite size effects in SA and explain them by studying the BP behavior on single instances of finite size. Finally, in Sec.~\ref{sec:discussion} we summarize the results found, the general picture emerging from this work and possible future implications. We leave to the Supplemental Material the more technical computations.

\section{The planted coloring problem}
\label{sec:plantedCol}

In the present work, we focus our attention on the planted coloring problem, which was the first planted problem studied in detail \cite{krzakala2009hiding}. 
Let us start recalling briefly the coloring problem (both the random and the planted versions) and their phase transitions.

In the random $q$-coloring problem, a random graph of $N$ vertices and mean degree $c$ is given and we have to assign one among the $q$ available colors to each vertex in a way to avoid monochromatic edges, i.e.\ adjacent vertices with the same color.
In the thermodynamical limit $N\to\infty$, a typical random $q$-coloring problem undergoes several phase transitions \cite{mulet2002coloring,zdeborova2007phase}.
Before the so-called $\textit{dynamical}$ transition $c_d$, there are exponentially many (in $N$) solutions that all belong to a large single cluster. For $c_d<c<c_c$, with $c_c$ being the \textit{condensation} transition, the space of solutions splits in an exponential number of different clusters, whose number becomes sub-exponential for $c>c_c$.
A typical random graph is $q$-colorable up to the \textit{satisfiability} threshold $c_s$ and for $c>c_s$ there is no solution to a typical random $q$-coloring problem.
The actual values of the different transitions connectivities, $c_d \le c_c \le c_s$, depend on $q$ and also the nature of the phase transition does: for $q<4$ the dynamical and condensation transition coincide, leading to a continuous phase transition, while for $q\ge 4$ the transition is discontinuous. And this has consequences also on the corresponding planted model.

The planted coloring problem has been introduced in Ref.~\cite{krzakala2009hiding} and consists in first assigning a random $q$-coloring to the $N$ vertices (this is the planted solution) and then building the random graphs of mean degree $c$ by adding $M=cN/2$ edges connecting randomly chosen pairs of vertices of different colors.
The inference problem corresponds to the recovery of the signal (the planted solution) given the random graph $G$ built as explained above. So the mean degree $c$ of the random graph can be seen as a sort of signal to noise ratio: for small $c$ the graph $G$ has exponentially many solutions (colorings) and it is impossible to identify the planted one, while for very large $c$ the planted coloring is only compatible with the graph $G$ and can potentially be identified.

To achieve a more detailed description of the phase transitions taking place in the planted coloring problem, we need to discuss the solutions to the BP equations associated to the problem. Calling $\psi^{i\to j}_s$ the probability that the vertex $i$ takes the color $s$ in absence of the link between $i$ and $j$, the BP equations read \cite{zdeborova2007phase}
\begin{equation}
    \psi^{i\to j}_s=\frac{1}{Z^{i\to j}}\prod_{k\in\partial i \setminus j} (1-\psi^{k\to i}_s)
\end{equation}
where $Z^{i\to j}$ is a normalization imposing $\sum_s \psi^{i\to j}_s=1$ and $\partial i \setminus j$ indicates all the neighbours of $i$ with the exclusion of $j$.
The BP equations are usually solved iteratively and at convergence provide the marginal probability that variable $i$ takes color $s$ via
\begin{equation}
    \psi^{i}_s=\frac{1}{Z^{i}}\prod_{k\in\partial i} (1-\psi^{k\to i}_s),
    \label{eq:BP}
\end{equation}
where $Z^{i}$ is again a normalization constant.

The fixed point marginals are mainly of two kinds \footnote{For more complicated scenarios we address the reader to the thorough survey of phases in Ref.~\cite{ricci2019typology}}: (i) uniform over the $q$ colors ($\psi_s^{i}=\frac{1}{q}, \forall i,s$), identifying the paramagnetic (PM) phase; (ii) highly biased towards the planted solution, identifying the signal recovery or ferromagnetic (FM) phase. The stability analysis of the paramagnetic fixed point towards perturbations in the direction of the planted state reveals that the paramagnetic phase is locally stable for $c<\cKS=(q-1)^2$ \cite{krzakala2009hiding} (in that work the threshold was called $c_l$, but we prefer a name making evident the connection with the Kesten-Stigum bound in tree reconstruction). So for $c<\cKS$, we expect that BP, if initialized with no information about the signal, will converge to the uninformative PM fixed point and recovery via BP is impossible. Vice versa, for $c>\cKS$ any initial perturbation in the direction of the signal is amplified and BP will spontaneously converge to the informative FM fixed point, highly correlated with the signal. Thus $\cKS$ is the algorithmic threshold for partial signal recovery via BP.

The procedure described above is called `quiet planting' as it adds the planted solution (and a cluster surrounding it) without changing the structure of solutions to the random problem for $c<c_c$. Since in this region the planted cluster is one among the exponentially many others, finding it is impossible.
Instead, for $c>c_c$, the planted cluster of solutions dominates the total number of solutions and the planted solution can be detected in principle.
For this reasons $c_c=\cIT$ is called the \emph{information theoretical} threshold.

Whether the planted solution can be actually found above $\cIT$ depends on the nature of the phase transition \cite{zdeborova2016statistical}. In continuous phase transitions, we have $\cIT=\cKS$ and so, as soon as the planted cluster dominates the total number of solutions, BP can actually detect it. This happens in the planted $q$-coloring problem for $q<4$. On the contrary, if the transition is discontinuous, we have a \emph{hard} phase for $\cIT<c<\cKS$ where the planted solution is in principle detectable, but signal recovery via BP is impossible, due to the stability (and the strong attraction) of the PM fixed point.  This happens for $q>4$ in the planted $q$-coloring problem. The case $q=4$ has remained elusive for a while and was finally solved in \cite{ricci2019typology}.

In the present work we focus on the case $q=5$, that is an inference problem with a well defined hard phase. Threshold values for $q=5$ are $c_d=12.837(3)$, $c_c=\cIT=13.23(1)$ and $\cKS=16$ \cite{zdeborova2007phase, krzakala2009hiding}.
Despite a great effort to break this hard phase, no polynomial algorithm is known that can detect the planted solution for $c<\cKS$.

Let us finally discuss a delicate point.
By construction, the BP algorithm relies on the Replica Symmetric (RS) approximation, assuming that most of the solutions form a single and well-connected cluster \cite{mezard2009information}. Consequently, the BP fixed points can describe only RS phases, like the PM and FM phases discussed above. However, from the solution of the random model, we know that for $c>c_d$ the space of solutions splits into many different clusters: this is the so-called \emph{glassy} phase, as it induces a very strong slowing down in many algorithms, and its description requires to assume Replica Symmetry Breaking (RSB) \cite{mezard2001bethe}. In the planted graph, also the planted cluster starts to be locally stable at $c=c_d$, exactly as the glassy states. For $c_d<c<c_c$ the planted cluster shares the same properties as the random glassy states, while for $c>c_c$ its internal entropy becomes larger than the entropy of the glassy states, and it starts to be detectable. These glassy states, which are so important in the description of the random model, play no role in the BP behavior when it is applied to the planted case and the performance of the BP algorithm is not improved by taking into account the glassy nature of the hard phase \cite{antenucci2019glassy}.

We have already mentioned that BP is Bayes-optimal: it can find optimal solutions for $c>\cKS$ because it assumes the perfect knowledge of the model.
Indeed, Eq.~(\ref{eq:BP}) assumes that we know that there is a hard constraint imposing the absence of links between nodes of the same color in the construction of the graph around the planted solution.
It can be shown that, imposing the perfect knowledge of the model, there cannot exist an RSB phase, and the only possible transition is a usual ferromagnetic first order transition between PM and FM phases.
In statistical mechanics language, assuming the perfect knowledge of the model corresponds to be on the so-called \emph{Nishimori line} \cite{nishimori2001statistical}, where the RS assumption holds.
So, in the planted coloring problem, thanks to the perfect knowledge of the generative model, only PM and FM phases can dominate the measure over the solutions. The first order transition takes place at $\cIT$, where the free-energies of the two phases are equal, with the FM free-energy being lower for $c>\cIT$ (thus making the signal in principle detectable). The spinodal points are $c_d$ and $\cKS$, as the FM planted state appears at $c_d$ and the PM state becomes unstable at $\cKS$.

Leaving BP aside, and considering broader classes of algorithms it is very likely that one has to consider situations where algorithms do not satisfy the Nishimori condition: either because they have no perfect knowledge about the model, or because, even knowing the model, they go out of equilibrium, thus violating the Nishimori condition.
In all these cases, the glassy states --- so important in the random model --- should be taken again into account (even if they are irrelevant in the thermodynamic RS description of the planted model). We will see in the next sections that they will be particularly important to determine the behavior of MCMC-based solvers that do not assume the perfect knowledge of the model.
This is an ideal example to show that moving away from equilibrium statistical mechanics to out of equilibrium processes one can find a much richer physics that requires a more complex description.

\section{Simulated Annealing for the planted coloring problem}
\label{sec:SA}

In this section, we introduce the Simulated Annealing (SA) as a solver for the planted coloring problem, and we analyze its performance.
SA is an MC-based algorithm introduced in Ref.~\cite{kirkpatrick1983optimization} and widely used in many optimization problems, including the random coloring problem \cite{chams1987some,johnson1991optimization}.
In the context of the planted coloring problem, it was considered in Ref.~\cite{krzakala2009hiding}, where the authors stated it can find the planted solution in the whole region $c>\cKS$. We will show that this is not in agreement with our numerical results and propose a better conjecture.
 
Given a configuration $\bs$ of colors of the nodes, in a statistical mechanics approach, we associate to it an energy counting the number of monochromatic edges
\begin{equation}
    H(\bs)=\sum_{(i,j)\in E}\delta_{s_i,s_j}\;,
\end{equation}
where $E$ is the edge set of the graph.
We then introduce a temperature $T \equiv 1/\beta$ in the model and a Gibbs-Boltzmann-like probability measure on the configurations
\begin{equation}
    P_{GB}(\bs) \propto e^{-\beta H(\bs)}\;.
\end{equation}
In the $T=0$ limit, this probability measure becomes the uniform distribution over the set of solutions (proper colorings), which is always non-empty thanks to the presence of the planted solution $\bs^*$ with $H(\bs^*)=0$. Moreover the $T=0$ distribution coincides with the posterior $P(\bs|G)$ if the prior is uniform over all configurations. This indicates that the Bayes optimal temperature is $\TB=0$. One could then naively conclude that 
the best strategy for any inference algorithm should be the sampling of the $T=0$ distribution to detect $\bs^*$.
We will see that the optimal temperature for MC-based algorithms is different from the Bayes optimal one $\TB=0$.

The meaning of the distribution $P_{GB}(\bs)$ with $T>0$ is to relax the hard constraints into soft ones: configurations with monochromatic edges are admitted, but with a probability that is more suppressed the lower is the temperature.
We start our SA from a random configuration of colors and at sufficiently high $T$.
At each step of the SA algorithm we decrease the temperature by $dT$ and we do a Monte Carlo Sweep (MCS), that corresponds to the attempt to update the color of each of the $N$ variables following the usual Metropolis rule: in practice, we change the color $s_i$ into $s_i'$, chosen at random between the $q$ possible colors,  with probability 1 if in this way the number of monochromatic edges decreases, while we update it with probability $e^{-\beta n}$ if the number of monochromatic edges  increases by $n$ after the update. 

\begin{figure}[t]
\begin{center}
\includegraphics[width=\columnwidth]{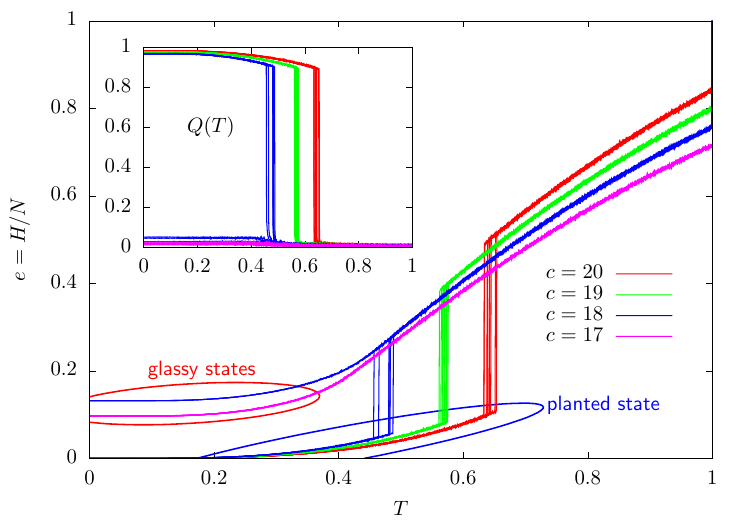}
\end{center}
\caption{The energy density as a function of the temperature in the Simulated Annealing for planted coloring with $q=5$, $N=10^5$ and $dT=10^{-7}$. For each value of the mean degree $c$, we show the results for 10 different samples. For $c\lesssim18$, SA does not find the planted solution, and get trapped in glassy states. Inset: overlap with the planted configuration as a function of the temperature (same parameters as in the main figure).}
\label{Fig:SA_variando_c}
\end{figure}

In Fig.~\ref{Fig:SA_variando_c} we show the intensive energy, $e=H/N$, of the configurations visited by SA with parameter $dT=10^{-7}$ in 10 samples of size $N=10^5$ of the planted coloring problem with $q=5$. Differences between samples are negligible (but at $c=18$), while the behavior of SA changes drastically with the mean degree $c$.

For $c>18$, SA finds a configuration highly correlated with the planted solution $\bs^*$ on each sample. This is confirmed also by the data in the inset, showing the overlap between the configuration $\bs$ reached by SA and the planted solution $\bs^*$ defined as
\begin{equation}
    Q=\frac{\max_{\pi\in S_q}\sum_{i} \delta_{s^*_i,\pi(s_i)}/N-1/q}{1-1/q}\;,
\end{equation}
where $S_q$ is the group of permutations of $q$ elements.
The overlap defined in this way is null for a random guess, while $Q=1$ if there is a perfect recovery of the planted solution.

On the contrary, for $c<18$, SA is unable to reach the planted solution in any sample, and it gets trapped in glassy states, which are metastable states (local minima of higher energy), orthogonal to the planted state ($Q \simeq 0$). So the numerical data strongly suggest that $\cSA \simeq 18$ is the algorithmic threshold for SA. Indeed, exactly at $c=18$, we see in Fig.~\ref{Fig:SA_variando_c} that the behavior of SA is sample dependent: a typical behavior that is often observed at a spinodal point.

The strict inequality $\cKS<\cSA$ implies SA is not Bayes optimal and there exists a region, $c\in(\cKS,\cSA)$, where BP performs better than SA.

Let us comment briefly on the choice of $N$ and $dT$. We have studied several values of $N$ and $dT$, all providing data consistent with the claims above, as long as $dT$ is small enough. In App.~\ref{app:dTvsN} we present the data for different $N$ and $dT$ values, showing that for $c>\cKS$ the SA algorithm can find solutions with high probability if $dT=O(N^{-a})$ with $a\simeq 1$. This means that SA is a solving algorithm running in a time scaling roughly quadratically with the problem size $N$. We decided to show only data for $N=10^5$ since this value is large enough that fluctuations are strongly suppressed.

Data in Fig.~\ref{Fig:SA_variando_c} clearly show that a dynamical first order transition is taking place at the temperature where SA makes a jump to reach the planted cluster of solutions. Moreover when such a dynamical first order transition does not take place (i.e. for $c<\cSA$) the overlap remains null and SA stays very far from (actually orthogonal to) the planted solution.

\begin{figure}[t]
\begin{center}
\includegraphics[width=\columnwidth]{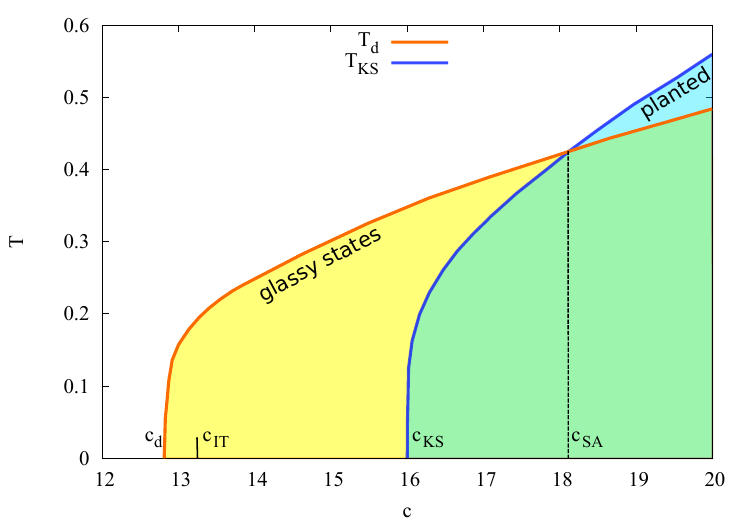}
\end{center}
\caption{Phase diagram in the $c-T$ plane for the planted-coloring problem with $q=5$.
$T_d(c)$ is the critical line below which glassy states appear, while below $\TKS(c)$ the paramagnetic solution becomes unstable toward the planted solution. The Information-Theoretical threshold $c_{IT}$ corresponds to $c_c$, while we claim that SA is not able to find the planted solution for $c<c_{SA}$. We can extract $c_{SA}$ as the connectivity at which $T_d(c)=\TKS(c)$. Both in the yellow and the green region, an MCMC algorithm gets trapped in glassy states, and the only region in which it can nucleate the planted solution is the blue one. On the other hand the BP algorithm finds the paramagnetic solution in the yellow region while it succeed to find the planted solution both in the blue and in the green one.}
\label{Fig:phaseDiagram_y1}
\end{figure}

To explain why SA is not able to find the planted solution for $c<\cSA$, we need to look to the extension to the finite temperature of the phase transitions at $T=0$. 
How the thresholds $c_d$ and $\cKS$ change in temperature was already computed in Ref.~\cite{krzakala2009hiding} (more details in the Supplemental Material): 
we report them with the names $T_d(c)$ and $\TKS(c)$ in Fig.~\ref{Fig:phaseDiagram_y1}. 
For $T<T_d$ glassy states appear while for $T<\TKS$ the paramagnetic solution is no more stable with respect to perturbations toward the planted solution. 

The SA protocol starts from very high temperatures and keeps cooling very slowly (ideally in an adiabatic way).
Which critical line, either $T_d$ or $\TKS$, it will find first depends on the $c$ value.
Defining the SA algorithmic threshold $\cSA$ as the crossing point of these two critical lines
\begin{equation}
    T_d(\cSA) = \TKS(\cSA)\;,
\end{equation}
we conjecture the following scenario.
If $c>\cSA$, decreasing $T$ the paramagnetic phase becomes unstable with respect to the planted state and SA jumps into the planted cluster, thus achieving perfect recovery in the $T=0$ limit (the planted state is very stable at low temperatures and SA cannot leave it when glassy states appear below $T_d$).
Conversely, if $c<\cSA$, decreasing $T$ the paramagnet first becomes unstable toward the glassy states at $T_d$.
Remind that for $T>0$ the Nishimori condition is not satisfied and so glassy states are not only possible but likely to play a central role.
And indeed we observe SA falling inside glassy states for $c<\cSA$.
The barriers to escaping these glassy states are extensive, and SA cannot jump anymore to the planted solution, even if the temperature is decreased below $\TKS$: once the dynamics is trapped by a glassy state, it will stay there for a time that is exponential in the size of the system. 

Based on the arguments above the conjecture that $\cSA$ is the algorithmic threshold for SA is very plausible, on top of being fully supported by numerical observations.

\section{Replicated Simulated Annealing}
\label{sec:RSA}

Having understood that the SA algorithmic threshold $\cSA$ is larger than the BP one $\cKS$ because of the existence of glassy states than can trap the SA algorithm, a natural question is whether another MC-based algorithm could do better than SA.
In this section, we study the Replicated Simulated Annealing (RSA) introduced in Ref.~\cite{baldassi2016unreasonable} and provide strong numerical evidence that its algorithmic threshold $\cRSA$ matches $\cKS$, thus being as good as the Bayes optimal BP.

The idea beyond RSA is to simulate $y$ replicas (copies) of the system coupled together with ferromagnetic coupling (i.e.\ replicas prefer to be in similar configurations).
The corresponding energy function is the following
\begin{equation}
H(\bs)=\sum_{a=1}^y\sum_{(i,j)\in E}\delta_{s_i^a,s_j^a}-\frac{\gamma}{2y} \sum_{a\neq b}\sum_{i=1}^N \delta_{s_i^a, s_i^b}\;,
\label{eq:RSA_H}
\end{equation}
where $\gamma$ controls the intensity of the ferromagnetic coupling between replicas, that we decided to scale by a factor $y$ in order to get a well defined limit when $y\to\infty$.
The idea of simulating different replicas is not new in the use of SA for inference problems \cite{hu2012phase}, and the correlation among replicas was also used to modify the underlying graph \cite{hu2012replica}. The novelty in the RSA is that the different replicas do not evolve independently but are coupled together. This is a practical way to probe states that dominate not the original free-energy, but a modified free-energy where the contribution of local entropy is enhanced.

From a practical point of view, RSA works exactly as SA: we start at a high enough temperature from random independent configurations for each replica.
Each MCS consists in the attempt to change the color of each variable in each replica, according to the energy in Eq.~(\ref{eq:RSA_H}).
After each MCS the temperature is decreased by $dT$. 
In all our numerical experiments we set $\gamma=1$ (a complete study of the optimal value of $\gamma$ could be performed, but it is beyond the scope of the present work).

The lowering of the temperature forces the system to look for lower energy configurations, both reducing the number of monochromatic edges in each replica and increasing the similarity between replicas.
In this way, RSA should converge towards low energy states of large entropy, as larger clusters of solutions can more easily accommodate $y\gg 1$ replicas.  
The reason why this should favor the planted cluster of solutions against the glassy states is that beyond $c_c$ the planted cluster has the largest entropy: in fact, glassy states are many, but the entropy of each glassy state is relatively small, such that their total entropy is not larger than the one of the states dominating the thermodynamics.
This picture is not limited to the present problem and generalizes to other inference problems where the signal competes with many more random ``glassy'' configurations.

\begin{figure}[t]
\begin{center}
\includegraphics[width=\columnwidth]{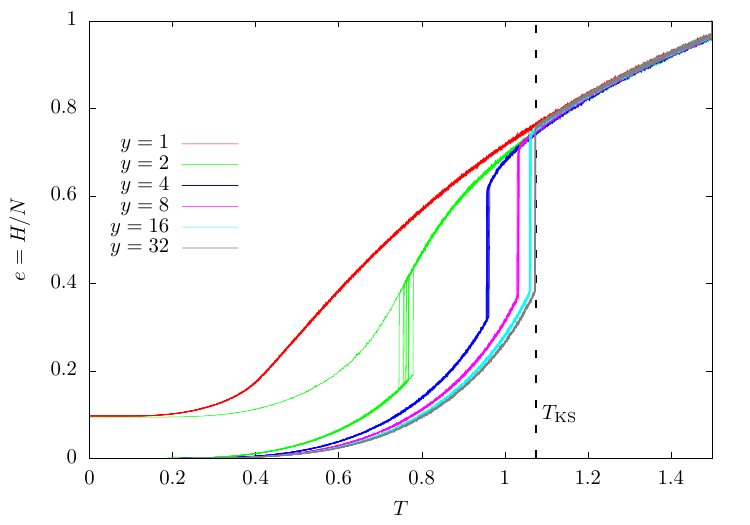}
\end{center}
\caption{The energy density as a function of the temperature in the Replicated Simulated Annealing for planted coloring with $q=5$, $N=10^5$, $dT=10^{-7}$ and $c=17$, changing the number $y$ of replicas. For each value of $y$, 10 different samples are simulated.
While for $y=1$ SA does not find the planted solution, as long as $y>1$ RSA finds it.
The vertical line represents $\TKS$ as found by the linearization of BP equations around the PM solution in the limit $y\to\infty$. 
}
\label{Fig:RSAc17}
\end{figure}

In Fig.~\ref{Fig:RSAc17} we show the behaviour of RSA for $N=10^5$ and $c=17$. We report data for only one value $dT=10^{-7}$, but in App.~\ref{app:dTvsN} we show that it is enough to have $dT=O(N^{-a})$ with $a\simeq 1$ to recover the signal. The curves for $y=1$ are the same shown in Fig.~\ref{Fig:SA_variando_c} reporting the inability of SA in finding the planted solution. Instead, RSA can find the planted solution as long as $y>1$. Surprising enough, already the smallest value $y=2$ seems very effective in the task of detecting the planted signal, with 9 over the 10 samples shown in the figure reaching a perfect recovery. The temperature where detection takes place, that corresponds to a jump in the energy, fluctuates a little bit among samples, while for larger values of $y$, the behavior of RSA is practically sample-independent and seems to follow an almost deterministic law. A possible explanation is that the PM state in the replicated model is becoming unstable at the temperature where the jump takes place: if this happens the RSA dynamics has no other option than flowing towards the planted solution, thus eventually detecting the signal in the $T=0$ limit. 

\begin{figure}[t]
\begin{center}
\includegraphics[width=\columnwidth]{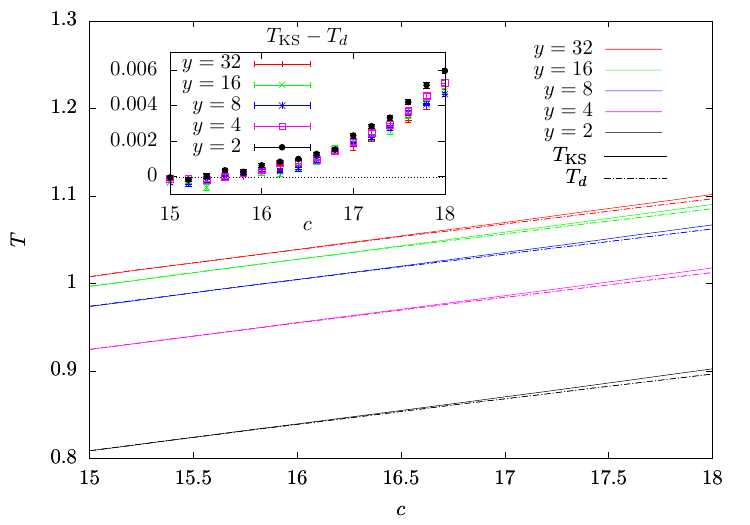}
\end{center}
\caption{$\TKS$ (full lines) and $T_d$ (dashes lines) as a function of $c$ in the replicated model with $y=2,4,8,16,32$. Data are averaged over 10 samples of size $N=10^6$ (which is large enough to avoid any significant finite size effects). Inset: the difference $\TKS-T_d$ is plotted to help locating the crossing point that defines the algorithmic threshold $\cRSA$, perfectly compatible with the BP threshold $\cKS=16$. Below this threshold the difference $\TKS-T_d$ is compatible with zero, thus making plausible the scenario where the temperature at which the paramagnetic state loses its local stability is the same in the planted and the random ensemble.}
\label{Fig:New_diagram}
\end{figure}

To test the above hypothesis we compute $\TKS(c)$ in the replicated planted model. Moreover, to test the conjecture of the previous section locating the algorithmic threshold $\cRSA$ for RSA, we compute also $T_d(c)$ in the replicated random model.
The details of these computations are presented in the Supplemental Material and the main results are summarized in Fig.~\ref{Fig:New_diagram} for $y=2,4,8,16,32$.
It is immediately evident that the critical lines for $y>1$ are located at much higher temperatures than those for the original model ($y=1$) shown in Fig.~\ref{Fig:phaseDiagram_y1}.
And these critical temperatures are indeed very close to the temperature where the jump takes place.
In particular, the vertical dashed line in Fig.~\ref{Fig:RSAc17} marks the value of $\TKS$ in the $y\gg 1$ limit, which turns out to be a very accurate analytical prediction for the jump location (i.e.\ the signal detection) in RSA run with $y\gg1$ and a very good approximation also for finite values of $y$.
We have seen that the sample to sample fluctuations in $\TKS$ are larger in the case of small $y$ and becomes almost null in the large $y$ limit, again in perfect agreement with the observation of the temperature of the jumps for RSA.
Moreover the critical lines $\TKS(c)$ and $T_d(c)$ shown in Fig.~\ref{Fig:New_diagram} for $y>1$ are extremely close. Indeed we report in the inset of Fig.~\ref{Fig:New_diagram} their difference to help the reader in locating their crossing-point that defines the algorithmic threshold $\cRSA$ (according to the conjecture put forward in the previous section).
Within the statistical uncertainties, the difference $\TKS-T_d$ does not depend on $y$ and becomes null at $\cRSA\simeq 16$, perfectly compatible with the algorithmic threshold $\cKS=16$. 
Thus RSA and the Bayes optimal BP algorithm are observed to share the same (conjecturally optimal) algorithmic threshold.
Please note that below $\cRSA$ the difference $\TKS-T_d$ is compatible with zero: for these connectivities the temperature at which the paramagnetic state loses its local stability is the same in the planted and the random ensemble.
As long as $y>1$, changing the number of replicas $y$ does not seem to lead to improvements, as we also see directly from the RSA simulation (see Fig.~\ref{Fig:RSAc17}). The best choice is thus to use a small $y$ value, even $y=2$, because the time for running RSA scales as $O(y\cdot N)$.

Let us finally discuss an important difference between the original model and the replicated one. For $y>1$, in the random model, the transition towards the glassy states becomes continuous. So, while in the original model ($y=1$) glassy states at $T_d$ are already well-formed states and become dominant with a discontinuous transition, in the replicated model ($y>1$) the glassy states arise exactly at $T_d$ with a continuous transition (a plot showing how the order parameter changes introducing replicas is present in the Supplemental Material). This observation allows us to compute $T_d$ very easily for $y>1$ via the linearization of the BP equations in the random model, similarly to the computation of $\TKS$ in the planted model (more details in the Supplemental Material).
The consequences of the changing of the order of the transition on the existence of hard phases are discussed in the concluding paragraph.

\section{Finite size effects}
\label{sec:FSE}

\begin{figure}[t]
\begin{center}
\includegraphics[width=\columnwidth]{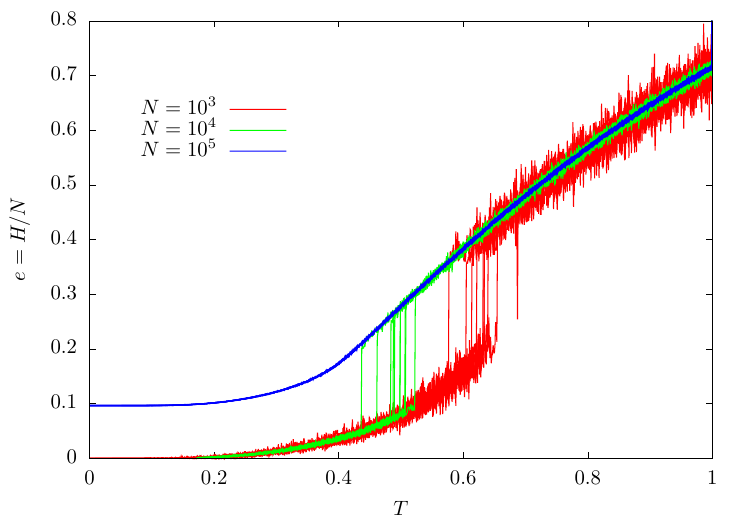}
\end{center}
\caption{The energy density as a function of the temperature during Simulated Annealing for planted coloring with $q=5$, $c=17$, $dT=10^{-7}$, and different problem sizes. For each value of $N$, 10 different samples are simulated. For $N=10^5$ SA does not find the planted solution, while it always does for smaller sizes.}
\label{Fig:SA_variando_N}
\end{figure}

Finite size effects are sizeable only close to the critical point.
As a typical example, we show in Fig.~\ref{Fig:SA_variando_N} the behaviour of SA at connectivity $c=17<\cSA$ for sizes $N=10^3,10^4,10^5$.
At first sight, the behavior of SA seems strange enough: it can find the planted solution for $N\leq 10^4$, even if the PM state should be stable given that we are running SA with $c<\cSA$.
As long as the PM is locally stable, the barrier one has to cross to leave it is $O(N)$ and thus the corresponding timescale should be $O(\exp(N))$, definitely a huge number for the sizes $N=10^3,10^4$ shown in Fig.~\ref{Fig:SA_variando_N}.

How can we explain this observation? Can we achieve an analytical prediction, at least qualitatively, of the above finite size effects?
A clear and simple answer comes from the study of the critical values $\TKS$ and $T_d$ on \emph{samples of a given finite size}.
Indeed both critical values can be computed by running BP on a given sample:
$\TKS$ is found by analyzing the stability of the PM state in the planted model via the linearization of BP equations around the paramagnetic fixed point;
$T_d$ is found by running BP on the random model initialized in a planted glassy state at a given temperature and looking at the temperature at which the glassy state becomes unstable towards the paramagnet.
More details are given in App.~\ref{app:Td}.

\begin{figure}[t]
\begin{center}
\includegraphics[width=\columnwidth]{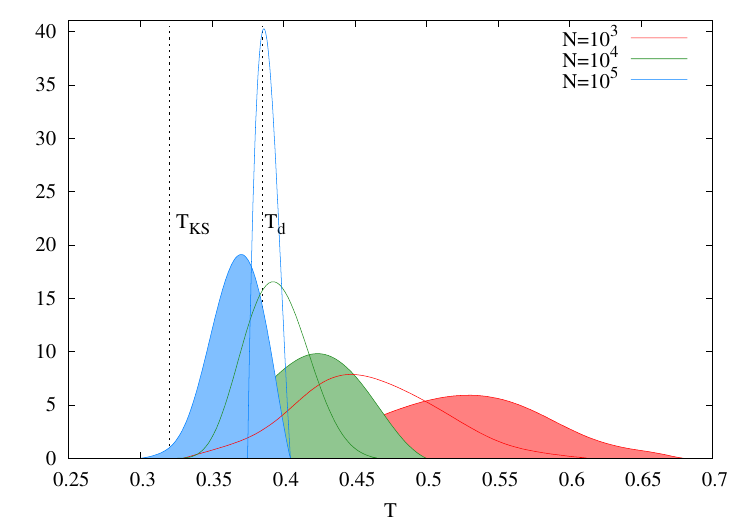}
\end{center}
\caption{Histograms of $\TKS$ (filled curves) and $T_d$ (empty curves), obtained by running linearized BP equations on given instances of different sizes at $c=17$ (more details on the computation in the text). Ordering values of $\TKS$ and $T_d$ at a given $N$ and pairing them following their order, for $N=10^3, 10^4$, $\max(\TKS,T_d)=\TKS$ for almost each pair of temperatures, while for $N=10^5$ it always happens that $\max(\TKS,T_d)=T_{d}$ (and this is the reason why SA is not able to reach the planted state for $N=10^5$). Vertical dashed lines represent $\TKS$ and $T_d$ in the thermodynamic limit.}
\label{Fig:Isto_Tks_y1}
\end{figure}

We cannot compare directly $\TKS$ and $T_d$ for a given sample, because, as explained exhaustively in the Supplemental Material, we are looking at the ensemble of graphs planted at $T=0$ to compute $\TKS$, and to the ensemble of graphs planted at $T>0$ to find $T_d$. 
However, to compare the two, we can reasonably assume that the finite size is effectively changing the mean degree $c$. $T_d$ and $\TKS$ being both monotonic functions of $c$, we can assume that the sample that has the highest $T_d$ will also have the highest $\TKS$. Following the same reasoning, we order values of $\TKS$ and $T_d$ and pair them following their order.
For each pair, we can now look to the $\max(\TKS,T_d)$, because we conjecture that the highest of the two is the one determining the fate of SA in the search for the planted signal.
In Fig.~\ref{Fig:Isto_Tks_y1} we show the histograms of $\TKS$ (filled curves) and $T_d$ (empty curves) for different sizes at $c=17$. 
The data are thus in perfect agreement with our claim that SA can find the planted state only if $\max(\TKS,T_d)=\TKS$ and this happens with a high probability for $N=10^3$ and $10^4$, while we never observed it for $N=10^5$.

The reason beyond this unexpected behavior is due to the very different finite size effects in $\TKS$ and $T_d$: while the latter deviates slightly from the value it takes in the thermodynamic limit, we observe in $\TKS$ substantial deviations, very biased towards larger temperatures. This in turn destabilizes the PM state at high temperatures in samples of finite size allowing for signal detection (a very favorable effect of finite size effects!).

\begin{figure}
    \centering
    \includegraphics[width=0.8\columnwidth]{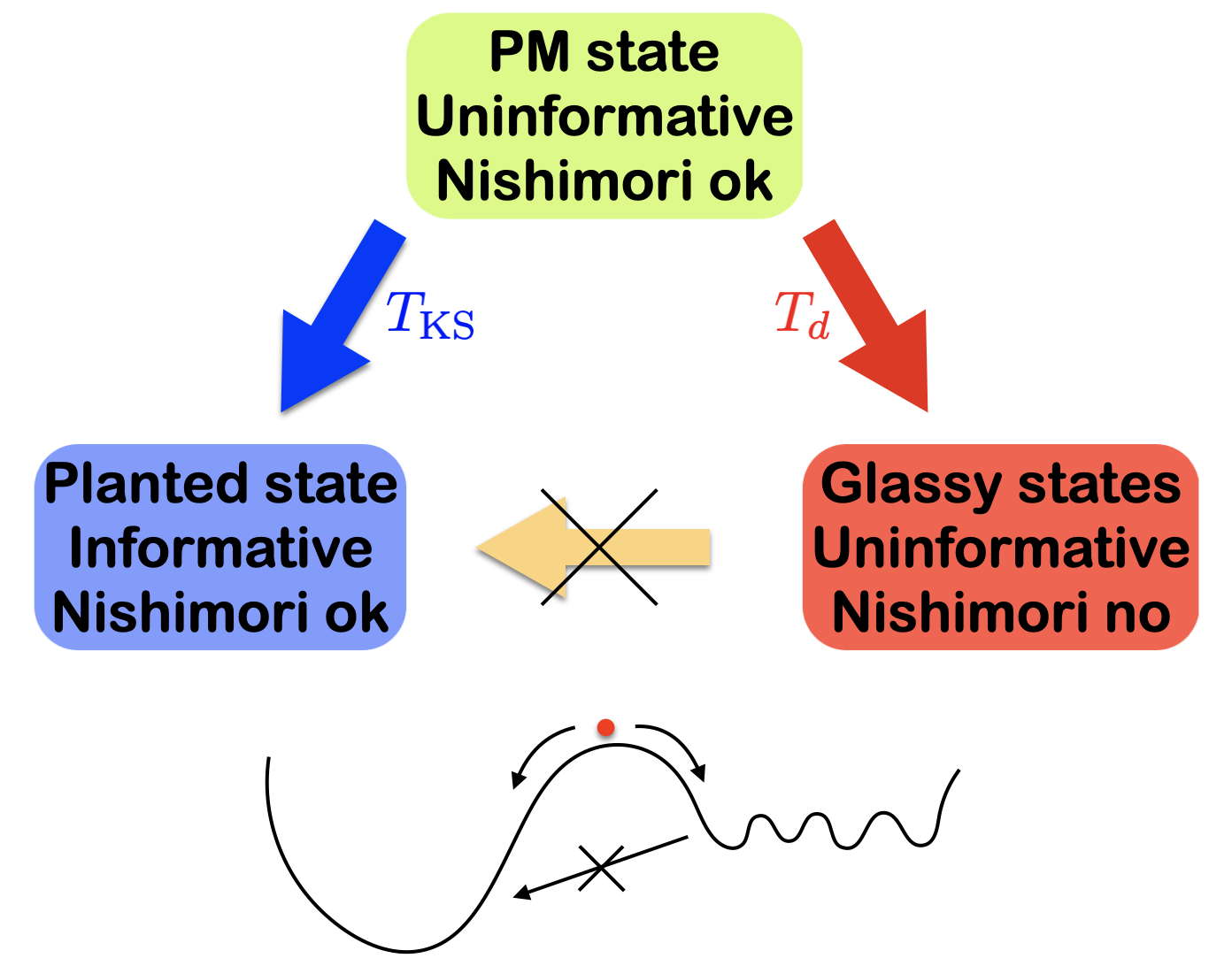}
    \caption{A schematic picture of the general scenario faced by an annealing process looking for the planted signal. Decreasing the temperature the fate of the annealing algorithm is determined by which critical temperature is met first between $\TKS$ and $T_d$ (as explained in the text). The lower diagram should remind us that once the system gets trapped in glassy states it cannot reach anymore the planted signal due to the large barriers.}
    \label{fig:schematic}
\end{figure}

\section{Discussion and perspectives}
\label{sec:discussion}

In this work we have analyzed the limits and the performances of simple, but efficient, algorithms based on MCMC (i.e.\ SA and RSA) in solving the planted coloring problem on sparse random graphs. This problem is a well-understood inference problem (in the teacher-student scenario), showing an information-to-computational gap: the planted solution can in principle be detected for $c>\cIT$, but any known polynomial algorithm can find it only for $c>\cAlg\ge\cKS>\cIT$.

For planted problems defined on random graphs and for which the generative model is known --- two conditions not always easy to meet --- the BP algorithm is Bayes-optimal and can find the planted solution as soon as $c>\cKS$. The BP algorithm works at the same temperature used in the planting process ($T=0$ in the present model) and for this reason, it satisfies the Nishimori condition that keeps the algorithm away from glassy states: indeed in glassy states, the replica symmetry is spontaneously broken, while the Nishimori condition implies the replica symmetry must be preserved.
As a consequence, BP is the optimal one among a large class of message passing algorithms: as shown in Ref.~\cite{antenucci2019glassy} generalizing BP to include RSB effects leads to worst algorithmic performances (probably due to the presence of the uninformative glassy states that we have seen trapping the MCMC algorithms).

Willing to understand the more general situation where algorithms are run with a temperature $T$ different from the planting temperature used in the problem generation ($T>T_p=0$ in our case), we need to abandon the Nishimori condition.
In this situation, the glassy states, which are uninformative from the point of view of the inference problem, can play a key role, influencing negatively the performances of the algorithms.
We have shown in the present work that algorithms based on MCMC, like SA, can indeed be trapped by glassy states and fail to detect the signal, even in the region where the FM state has lower energy with respect to the PM state.

Decreasing the temperature in a SA (or RSA) simulation the system faces a critical choice, which is schematically depicted in Fig.~\ref{fig:schematic}: when it leaves the PM state it can either flow to the FM state (and thus detect the planted signal) or get trapped in glassy states (which are uncorrelated to the planted signal). In the latter case, the FM state becomes unreachable, even if it has lower energy because the dynamics in the glassy states is extremely slow and the barriers (both energetic and entropic \cite{bellitti2021entropic}) make paths leaving the glassy states extremely improbable.

The above observations allow us to make a conjecture on the location of the algorithmic threshold for MCMC-based algorithms. In the literature, there is not a clear indication on which is the correct threshold for these algorithms: while some works claim that MCMC solvers can reach the BP threshold \cite{krzakala2009hiding}, some other references suggest that indeed glassy states do influence negatively the performances of the non-Bayes optimal algorithms \cite{sompolinsky1990learning,mannelli2020marvels,mannelli2019passed,sarao2019afraid,antenucci2019glassy}, but do not provide any estimate of the actual algorithmic threshold.

In this work, we conjecture that the fate of an MCMC-based annealing algorithm depends on which critical line it meets before. If the PM spinodal line $\TKS(c)$ is met first, the MCMC evolution spontaneously leaves the PM state heading towards the FM states (the only existing before the appearance of glassy states) and detecting the planted signal. On the contrary, if the dynamical critical line $T_d(c)$ is met first, glassy states become the most attractive for the MCMC evolution and the system gets trapped in glassy states forever.
We have seen that the above conjecture predicts with high accuracy the algorithmic thresholds measured in numerical simulations for both SA and RSA.
For SA we find that the algorithmic threshold is suboptimal ($\cSA>\cKS$), thus correcting previous claims of optimality \cite{krzakala2009hiding}.
For RSA, instead, we find an optimal algorithmic threshold ($\cRSA\simeq\cKS$).

The optimality of RSA is a very good news: being an MCMC-based algorithm, it is extremely robust and can be used with confidence also in more structured problems, like those coming from realistic applications, where message passing algorithms are likely to fail.
The RSA algorithm has been introduced in Ref.~\cite{baldassi2016unreasonable} as a smart way to sample configurations with large local entropy.
To the best of our knowledge, it was never used before in inference problems. We have adopted it with the following rationale: according to the scenario depicted above, the inference process can be seen as a competition between the single FM planted state and the exponentially many glassy states, the two competitors having similar chances of attracting the dynamics close to the algorithmic threshold. Given that the glassy states are many, we expect the basin of attraction of a \emph{single} glassy state to be much smaller than the one of the FM planted state. 
For this reason, in the replicated system the $y$ copies are more likely to be attracted by the FM state having a larger basin than by any single glassy state (while being distributed over several glassy states is not convenient because of the coupling among replicas).

In other words, the improvement of RSA over SA is due to the planted state being different from glassy states (and this is often the case in inference problems), but it can not be extended straightforwardly to other types of problems.
For example, in a recent paper we compared the performances of SA and RSA for an optimization problem, the largest independent set problem on sparse random graphs \cite{angelini2019monte}, and in that case, we did not find a great improvement in the replicated version.
However, in hard optimization problems, one is often trying to find a solution that lies inside a glassy state (e.g.\ in the clustered phase of constraint satisfaction problems \cite{krzakala2007gibbs,zdeborova2007phase,montanari2008clusters}). Thus, if the competition is taking place between glassy states of similar internal entropy, the replicated version of the algorithms does not lead to any great improvement (as we saw in the largest independent set problem \cite{angelini2019monte}).
So, it is worth stressing that the very good performances of RSA in inference problems~\footnote{In addition to the results shown in the present work about the planted random graph coloring, we have preliminary results also for the planted bicoloring of random hyper-graphs supporting our statements.} is mainly due to the internal entropy of the planted state being larger than the one of each glassy state, and this difference is general, being the planted state the one with the lowest energy and being the basin of attraction larger for lower energy states (see e.g.\ the discussion in Appendix A of Ref.~\cite{bellitti2021entropic}).

We have studied annealing algorithms as they represent the safer way to test a broad range of temperatures: any good practitioner would start running SA if she knows anything about the model to be optimized, and the output of SA would already provide a rough but useful indication of the critical temperature region to be better explored.
However, once one gets a more detailed picture of the energy landscape (as we did here for the planted coloring problem), the natural question is "What is the optimal temperature to run MCMC, eventually the replicated one?".
The answer is clear from Figs.~\ref{Fig:SA_variando_c} and \ref{Fig:RSAc17}: the optimal temperature for MCMC is the one where the jump to the planted state takes place (let us call it $\Tjump$).
Indeed for $T>\Tjump$ MCMC is stacked in the PM phase, while decreasing the temperature the dynamics becomes slower and the time for doing the jump increases (and even eventually diverges if glassy states with extensive barriers appear).

Noticing that in all cases where the signal can be detected by MCMC-based algorithms we have $\Tjump>0$, we can straightforwardly conclude the optimal MCMC temperature is not the Bayes optimal one.
A similar conclusion has been achieved recently for MCMC-based algorithms solving the planted clique problem \cite{angelini2021mismatching,angelini2018parallel} and in the rank-one matrix estimation problem \cite{pourkamali2021mismatched}.
So, it seems a general statement that, once we consider out of equilibrium processes solving an inference problem, the Bayes optimal temperature plays a marginal or null role and the optimal temperature is typically higher (larger thermal fluctuations enhance the exploration of the configuration space).
This is reminiscent of what has been found for Langevin dynamics in another inference problem, the spiked matrix-tensor model: in that case, the Langevin algorithm is sub-optimal with respect to message passing algorithms, due to the presence of glassy states, and it is computationally advantageous for it to mismatch the parameters to find the planted solution of the problem \cite{mannelli2019passed,sarao2019afraid,mannelli2020marvels}.

The mismatch between the optimal MCMC temperature and the Bayes-optimal temperature is a good news for all those situations when the generative model is not known and matching the planting temperature would be simply impossible by ignorance.
So, having robust algorithms that can work even without knowing exactly the properties of the generative model is extremely useful, and the only requirement is to have an idea of the optimal temperature at which the algorithm should be run (the jump temperature in the present case).

We have attempted to get an analytical estimation of the jump temperature $\Tjump$, by checking the stability of the PM state in the replicated model (the details of the computation are in App.~\ref{app:KS}). However, the replicated model has many short loops and the Bethe approximation is no longer exact (one should consider super-variables involving the $y$ replicas and taking $q^y$ values, but we leave this computation for future work). Luckily enough in the large $y$ limit, the interaction among replicas becomes weak (we scaled it by $1/y$ to get a well-defined limit for $y\to\infty$) and the Bethe approximation is again valid on a single graph. In the $y\gg1$ limit we can thus obtain an estimate of $\Tjump$ (it is reported in Fig.~\ref{Fig:RSAc17} with a vertical dashed line) that compares extremely well with numerics even for smaller values of $y$. This is, in our opinion, a very relevant achievement as predicting the dynamical behavior of out of equilibrium algorithms is typically very challenging and the cases where this can be put in connection to the thermodynamical phase diagrams (as those shown in Figs.~\ref{Fig:phaseDiagram_y1} and \ref{Fig:New_diagram}) are very few, especially for sparse models.

\begin{figure}
    \centering
    \includegraphics[width=\columnwidth]{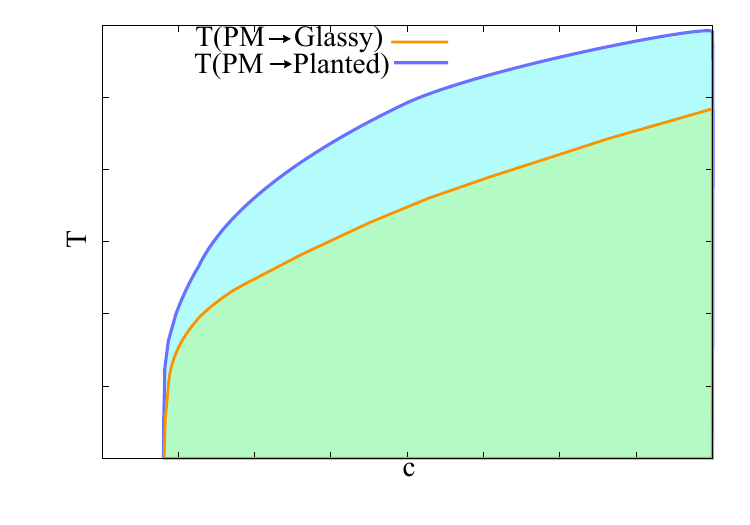}
    \includegraphics[width=\columnwidth]{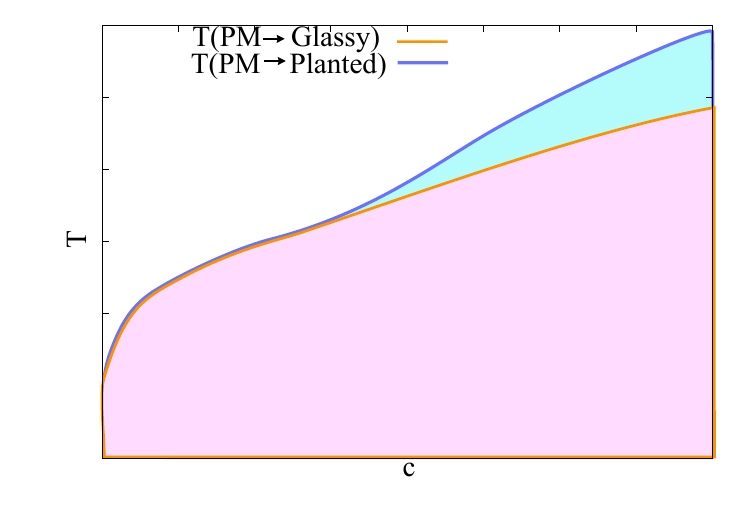}
    \caption{Qualitative phase diagram for a ``standard'' model with a continuous transition (upper panel) and for the replicated coloring model (lower panel). While for a standard continuous model the temperature at which the PM state becomes unstable towards the planted state is always higher than the temperature of the instability towards a random glassy state, for the replicated model below a certain connectivity the two temperatures coincide. As in Fig.~\ref{Fig:phaseDiagram_cont} in blue regions both an MCMC algorithm and BP can nucleate the planted solution while in the green one MCMC algorithms get trapped in glassy states while BP is able to find the planted one. In the diagram of the replicated model a new previously unseen region, coloured pink, appears where both an MCMC algorithm and BP are trapped by a glassy state.}
    \label{Fig:phaseDiagram_cont}
\end{figure}

A final comment is about the nature of the transitions in the replicated system. We already pointed out that the transition from the PM state towards the glassy states changes its nature adding replicas, passing from a discontinuous transition for $y=1$ to a continuous transition for $y>1$. This could sound strange and unexpected, however this phenomenon is reminiscent of what happens in other disordered systems: Let us consider the formulation of the problem where the $y$ coupled variables on a single site form a super-variable taking $q^y$ values. The super-variable is subject to a local field that favours configurations where the $y$ replicas are similar. It is well known that disordered models undergoing a discontinuous Replica Symmetry Breaking transition (e.g.\ the famous $p$-spin model \cite{crisanti1992sphericalp}) can change their nature to a continuous transition in presence of an external field. In the case of the $y$ replicas, the change in the nature of the transition is caused by a self-induced internal field.
This change in the type of transition should lead us to rethink the whole concept of the hard phase in inference problems, which is fundamentally grounded on the discontinuous nature of the phase transition. Indeed until now, any known model whose random version shows a discontinuous transition, has a hard phase in its planted version, while any model that shows a continuous transition in its random version, does not have a hard phase in its planted version. The phase diagram of a model with a discontinuous transition has the same form as the one we have already shown in Fig.~\ref{Fig:phaseDiagram_y1} for the coloring problem. In the upper panel in Fig.~\ref{Fig:phaseDiagram_cont} we show the typical phase diagram for a model with a continuous transition: the PM state loses stability as long as the glassy states or the planted state become locally stable respectively in the random graph or in the planted graph. The temperature at which the PM state becomes unstable towards the planted state is always higher than the temperature of the instability towards a random glassy state in a planted graph: as a result, there is no hard phase, because it is possible to recover the planted state as long as it becomes locally stable, both for a BP or for an MCMC algorithm. The phase diagram for the replicated model is instead different as shown in the lower panel of Fig.~\ref{Fig:phaseDiagram_cont}: below a certain connectivity (that in the case of the coloring problem we found numerically to coincide with $\cKS$) the temperature at which the PM state becomes unstable towards the planted state coincides with the one at which it becomes unstable towards the glassy states (in other words the two critical lines \emph{merge}). Moreover, as long as the glassy states become locally stable, both an MCMC algorithm \textit{and} the BP algorithm get trapped by them and do not find the planted state. In this situation the hard phase is present but its nature is different from the one arising in discontinuous models: while for discontinuous models the hard phase is caused by the local stability of the PM state and the planted state is retrievable as soon as the PM state becomes locally unstable, in the replicated model the continuous transition destabilizing the PM state may lead to the concurrent formation of glassy states (for $c < \cKS$ in our model) that trap the algorithms and make the planted state not retrievable.

At this point, one could ask why glassy states are not a problem for BP in the non-replicated models, both discontinuous and continuous (in the green region of both Figs.~\ref{Fig:phaseDiagram_y1} and \ref{Fig:phaseDiagram_cont} glassy states are stable but BP does not ``see'' them) while they trap the BP algorithm in the replicated case. In the non-replicated models, the number of glassy states goes to infinity in the thermodynamic limit, and their basin of attraction is very narrow: BP randomly initialized is not able to enter any of them. The introduction of replicas changes the free-energy of the model, performing a sort of weighted average over the states, reducing the number of glassy states and enlarging their basin of attraction (for a pictorial representation of this phenomenon see Fig.~1 of Ref.~\cite{baldassi2016unreasonable}): in this way they become accessible by the BP algorithm. This phenomenon should depend on the parameters $y$ and $\gamma$ and we leave its complete characterization for future works~\footnote{In previous works \cite{baldassi2016unreasonable} BP in the replicated model was found to converge to the planted state only if helped by \emph{focusing}, that is an effective slow change in $\gamma$ or in $T$. We have observed that standard BP does converge to the planted state if the right choice is made for $\gamma$ and $T$ to be in the blue region of the corresponding phase diagram.}. Please note that the PM state, which is trapping BP in the hard phase in the non-replicated model, can be viewed as the white average over all the underlying glassy states \cite{antenucci2019glassy}. When replicas are added, the role of PM trapping BP is taken by some new glassy states that can be viewed as a \textit{weighted} average over the glassy states of the non-replicated original model.

In conclusion, what is the most robust, eventually optimal, and easy-to-use algorithm to solve hard inference problems?
We have seen that for the coloring problem, RSA with few replicas, say $y=O(10)$, is the best option.
It is more robust than BP and it can be applied essentially to any kind of graph. 
Its running time is $O(y N)$, so only a factor $y$ slower than the simpler SA, which is the first option for any practitioner.
Maybe parallel tempering (PT) could be better than RSA, but it is more difficult to use and requires a pre-processing operation where the temperature schedule has to be optimized.
Moreover, RSA can be restricted to a small temperature range around $\TKS$, thus reducing the running times, while PT requires many more temperatures to be run efficiently.

Let us also comment that RSA can be viewed as a standard annealing process, that is a dynamical process trying to minimize the energy function, but using an effective Hamiltonian or cost function where configurations are \emph{reweighted}.
In the original work introducing RSA the motivation was to sample configurations of large local entropy \cite{baldassi2016unreasonable}.
However the idea can be generalized and we believe it is really worth exploring further the possibility of increasing the chances of reaching the planted signal by modifying the energy landscape. This has been shown useful in random constraint satisfaction problems, where the reweighting of solutions can modify the critical lines and enhance the chance of finding a solution \cite{budzynski2019biased,cavaliere2021optimization}.
Also in the tensor PCA problem, the modification of the energy landscape achieved via the use of many replicas can lead to sensible improvement in signal detection \cite{biroli2020iron}.

Having found an MCMC-based algorithm (RSA) that has the same algorithmic threshold $\cKS$ as the Bayes-optimal BP, some questions naturally arise (that we leave for future works).

Does the RSA optimality hold in general? A simple way to paraphrase this question is: Can one demonstrate that the equivalence
$$T_{KS} (c_{KS}, y) = T_d (c_{KS} , y)$$ is valid for all the planted inference problems whenever $y>1$ or at least for $y$ large enough?
And if yes, why? Can we get an improved RSA by optimizing the coupling among replicas?
We believe the answer to these questions will come from studying numerically more inference problems and analytically the $y\to\infty$ limit.

Does there exist an even better MCMC-based algorithm that can work below $\cKS$, i.e.\ breaking the hard phase?
The most immediate answer would be negative, given that BP (a Bayes-optimal algorithm) stops working at $\cKS$.
Moreover recent works connect the performance of a broad class of algorithms, called \emph{stable algorithms}, to the structure of the configurational space, suggesting that the clustering of solutions in the hard phase is an insurmountable obstacle for stable algorithms \cite{gamarnik2021overlap}.

However, we notice that MCMC-based algorithms do not belong to the class of stable algorithms (at least if run for a long enough time), so there is a hope they can work efficiently even in a clustered hard phase, where message-passing algorithms are deemed to fail.
Moreover smarter MCMC-based algorithms, like RSA or PT, actually work in an extended space (in replicas the former and in temperature the latter), and in this extended space, the clustering of the measure may be less important.
In support of this conjecture, we have shown in the present work how the nature of the phase transition changes in the replicated problem.

A last comment is on the scaling of the running times to find solutions in the regime where the algorithms we have studied do work, i.e.\ above $\cKS$.
Given that we focused on sparse problems, each iteration step of an algorithm is linear in $N$. However, the number of iterations can scale with the system size $N$.  
For message-passing algorithms, the number of iterations scales as a small power of $N$ or a large power of $\log(N)$, while for MCMC-based algorithms the number of iterations scales roughly linear in $N$ (see App.~\ref{app:dTvsN}).
This means we are already exploring the region of timescales growing with the system size, where analytical tools are at present unable to provide any detailed insight.
Without reaching the timescales growing exponentially in $N$, which are practically useless for any real-world problem, we believe the possibility of developing polynomial in $N$ solving algorithms based on MCMC working below $\cKS$ in the hard phase is to be seriously taken into account.

\section{Acknowledgments}

We thank Arianna Rampini and Leonardo Cortelli for the interesting discussions. We are also grateful to the reviewers for their insightful comments. This research has been supported by the European Research Council under the European Unions Horizon 2020 research and innovation program (grant No.~694925 -- Lotglassy, G Parisi) and by ICSC – Centro Nazionale di Ricerca in High Performance Computing, Big Data and Quantum Computing, funded by European Union – NextGenerationEU.

\appendix

\section{Summary of critical temperatures, their physical meaning and their connection with previous definitions}

We define $\TKS(c)$ as the spinodal temperature at which the paramagnetic state develops an instability towards the planted solution. This temperature was already computed in Ref.~\cite{krzakala2009hiding} under the name $T_2$ looking at the divergence of the ferromagnetic susceptibility. Its exact value is
$$
\TKS(c)=-\frac{1}{\log\[[\frac{c-(q-1)^2}{q-1+c}\]]}\;.
$$
Such a spinodal line ends at zero temperature at $\cKS=(q-1)^2$ (previously called $c_l$ in Ref.~\cite{krzakala2009hiding}). As usual in planted problems, $\cKS$ corresponds also to the connectivity at which the paramagnetic state becomes unstable in a random graph, without the planted state: this is signaled by the divergence of the \emph{spin-glass} susceptibility. However, one should be careful because this coincidence is not true anymore when temperature is added: the temperature at which the paramagnetic state becomes unstable in the random coloring problem has been computed in Ref.~\cite{zdeborova2007phase} and it reads
$$
T_c(c)=-\frac{1}{\log\[[1-\frac{q}{\sqrt{c}+1}\]]}
$$
and it is always smaller than $\TKS(c)$ for $c>\cKS$.

The other temperature that is fundamental in our analysis is $T_d(c)$, which is defined as the spinodal temperature above which a glassy state in the random coloring problem is not stable anymore. Its behaviour is already shown in Ref.~\cite{krzakala2009hiding}. At $T=0$ this line ends in $c_d = 12.837(3)$ for $q=5$ \cite{zdeborova2007phase}. Analogously to $\cKS$, $c_d$ results to be also the spinodal connectivity below which a planted configuration is not stable anymore. Again, this coincidence is not true anymore when temperature is added: the spinodal temperature for the planted configuration is the one called $T_1$ in Ref.~\cite{krzakala2009hiding}, and $T_1(c)>T_d(c)$ holds for $c>c_d$.

One can generalize the above definitions for $T_d(c)$ and $T_1(c)$ by introducing $T_\text{sp}(T_\text{pl},c)$ as the spinodal temperature at which a state planted at temperature $T_\text{pl}$ becomes unstable. With this definition, we have $T_1(c)=T_\text{sp}(0,c)$ and $T_d(c)=T_\text{sp}(T,c)$.

\begin{figure}
    \centering
    \includegraphics[width=\columnwidth]{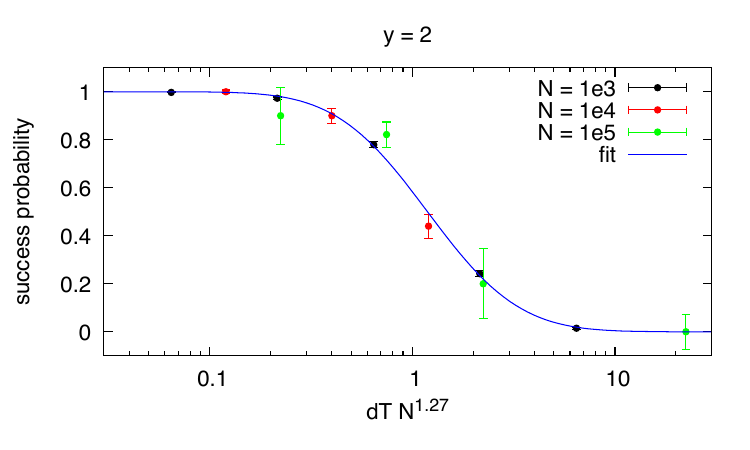}\\
    \includegraphics[width=\columnwidth]{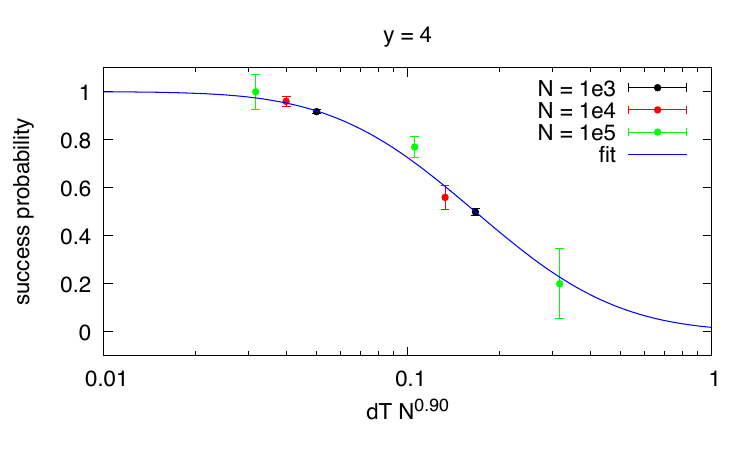}\\
    \includegraphics[width=\columnwidth]{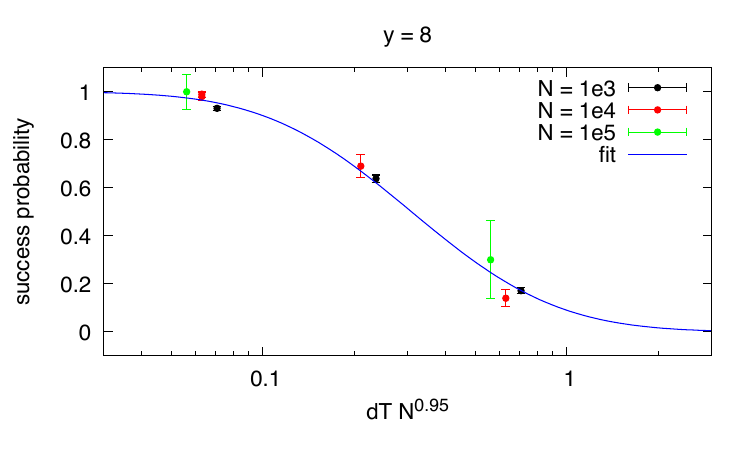}
    \caption{The success probability of finding the planted state by running RSA on $c=17$ planted coloring problems of size $N$ with a constant annealing rate $dT$ is a function of $dT N^a$ with a scaling exponent $a$ close to 1 (at least for $R=4,8$). This implies the running times of RSA are roughly quadratic in $N$. The fit function is $f(x)=\text{erfc}((x-x_0)/b)$.}
    \label{fig:RSAtimes}
\end{figure}

\section{Scaling of running times with system size}
\label{app:dTvsN}

In the main text, we have presented data only for $N=10^5$ and $dT=10^{-7}$, but it is important to study how small one has to set $dT$ in order to ensure that the annealing algorithm will converge to the solution with high probability. For any well-normalized cost function, we expect the starting temperature of the annealing process to be $O(1)$, i.e. not to scale with the system size $N$. As a consequence, the running time of the annealing process is $O(dT^{-1})$, and given that a single iteration in sparse models requires a time $O(N)$ the total running time is $O(N/dT)$. Let us discuss now how $dT$ should be scaled with $N$ in order to ensure convergence to the planted state.

We have run RSA for $c=17$ and SA for $c=20$ (in both cases we expect the algorithm to work if $dT$ is small enough).
We have considered sizes $N=10^3,10^4,10^5$ and several $dT$ values. From the data of the overlap presented in the inset of Fig.~\ref{Fig:SA_variando_c} it is clear that the condition $Q>0.5$ selects the successful runs. For each pair $(N,dT)$ we have measured the probability of a successful run.

In Fig.~\ref{fig:RSAtimes} we report data obtained by running RSA for $c=17$ with $y=2,4,8$, and a good scaling of the success probability can be achieved by plotting data as a function of $N dT^a$ with $a$ very close to 1. Using this renormalized variable, times for different sizes collapse and are well fitted by a function $t=\text{erfc}((N dT^a-x_0)/b)$ \footnote{This is the function expected if we consider the distribution of times-to-solution being a log-normal. If we put a threshold, the probability that the time is above the threshold is described by an error-function.}.
The best scaling exponents are $a \simeq 1.27$ for $y=2$, $a \simeq 0.9$ for $y=4$ and $a \simeq 0.95$ for $y=8$, but estimating the corresponding uncertainty is not easy as it depends on the finite size corrections to scaling which are not under control.
Nonetheless, all the estimated exponents are very close to 1 and this implies the RSA algorithm finds the planted solution in a time scaling roughly quadratically with the system size.

\begin{figure}
    \centering
    \includegraphics[width=\columnwidth]{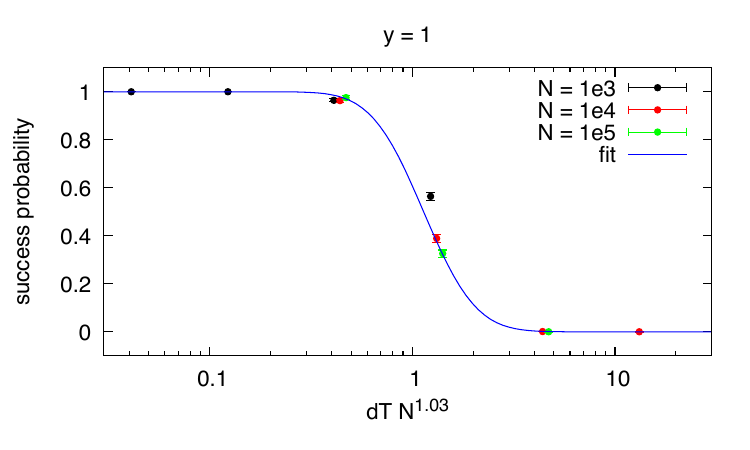}
    \caption{As in Fig.~\ref{fig:RSAtimes}, but for $c=20$ and running SA. Again the scaling $dT N^a$ with $a \approx 1$ is very good, although the smaller errors suggest the presence of finite size corrections to scaling.
    }
    \label{fig:SAtimes}
\end{figure}

\begin{figure}
    \centering
    \includegraphics[width=\columnwidth]{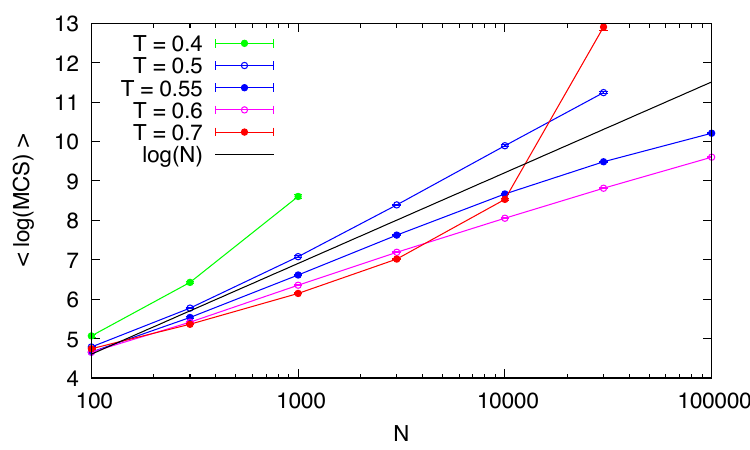}
    \caption{The mean of the logarithm of the number of Monte Carlo sweeps (MCS) needed by a Glauber dynamics at fixed temperature $T$ to find the planted state in $c=20$ coloring problems.
    }
    \label{fig:MCtimes}
\end{figure}

In Fig.~\ref{fig:SAtimes} we present the same scaling analysis for the performances of SA at $c=20$. Given the larger number of runs we have done, the errors on the success probabilities are much smaller and consequently make evident the need to use finite size corrections to scaling.
For this reason, we have computed the optimal scaling exponent, $a \simeq 1.03$, by collapsing data for $N=10^4,10^5$ assuming the data for $N=10^3$ are affected by some small corrections to scaling.

\begin{figure}
    \centering
    \includegraphics[width=\columnwidth]{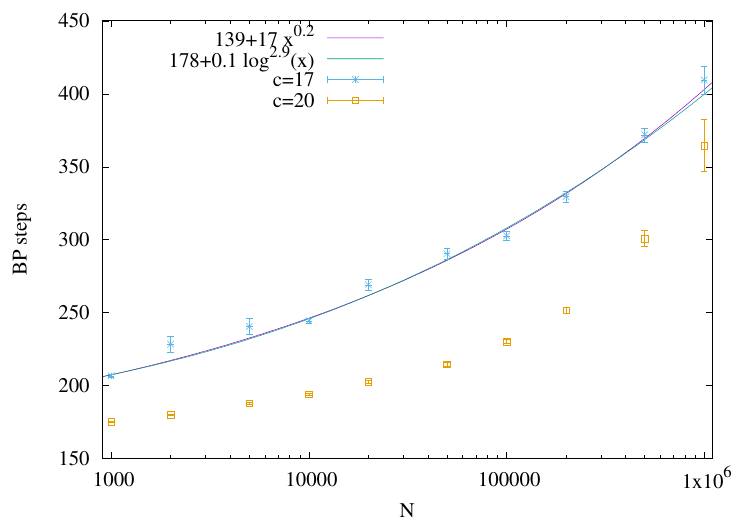}
    \caption{Number of iterations that the BP algorithm needs before convergence to the planted fixed point as a function of the size of the system $N$ for two values of the connectivities $c=17$ and $c=20$. The growth in the number of iterations can be fitted both by a power law $t(N)=a+bN^c$ with a very small exponent $c\simeq 0.2$, and with a logarithmic growth $t(N)=a+b \log^c(N)$ with $c\simeq 2.9$. The two curves are indistinguishable for the analyzed sizes and we cannot conclude which is the right scaling.}
    \label{fig:BPtimes}
\end{figure}

A simple numerical argument for the scaling $dT \approx N^{-1}$ in SA and RSA follows.
Assume that the Monte Carlo (Glauber) dynamics is able to detect the planted state only in the region $T_d<T<\TKS$ (i.e.\ the one coloured blue in Fig.~\ref{Fig:phaseDiagram_y1} and Fig.~\ref{Fig:phaseDiagram_cont}, right), that we call the \emph{retrieval region}.
The time spent in the retrieval region by annealing at a constant rate is $(\TKS-T_d)/dT$ and we need this time to scale like the time needed by an MCMC simulation run at any fixed temperature in the retrieval region to find the planted state.
We have measured the number of Monte Carlo sweeps (MCS) needed to nucleate the planted solution running a standard MCMC simulation at fixed temperature $T$ for $c=20$ and $y=1$. The results are presented in Fig.~\ref{fig:MCtimes}.
Straight data correspond to a power law growth with the problem size $N$ and the black line corresponds to a linear growth in $N$.
Several observations are in order.
The optimal time for small values of $N$ seems to be achieved for $T$ slightly above $\TKS \simeq 0.56$ and this is expected as setting the temperature to the largest possible value speeds up the MC dynamics and finite-size corrections make $\TKS$ larger (see Sec.~\ref{sec:FSE} on the finite size effects).
If $T$ is too large, eventually it becomes larger than $\TKS$ (that decreases towards its large $N$ limit) and leaves the retrieval region: when this happens the MC dynamics is no longer able to reach the planted state in a polynomial time and the corresponding time grows enormously with $N$ (see data for $T=0.7$).
If $T$ is below $T_d$ (see data for $T=0.4$) the time to find the planted state grows immediately very fast (likely super-polynomially).
Last, but not least, choosing the temperature in the retrieval region (better towards its upper end) the running times to find the planted state scale roughly linearly with the system size (the black line is a linear growth law for the ease of reading).
This last observation brings us to the conclusion that, although running an annealing algorithm is very comfortable because one does not need to spend time selecting the optimal running temperature (the annealing will visit the retrieval region wherever it is), it is not optimal, because all the time spent outside the retrieval region is actually useless.
The optimal and robust retrieval algorithm is running MC dynamics in the retrieval region. Since a single MCS takes a time of order $O(N)$, the time needed by MC dynamics to find the planted state in the retrieval region is again roughly quadratic.

As a comparison, in Fig.~\ref{fig:BPtimes} we show the number of iterations that the BP algorithm needs to converge to the fixed point corresponding to the planted state. Each single BP iteration takes a time of order $O(cN)$.
As shown, the growth of the number of iterations with the size of the system can be fitted both by a power law with a very small exponent and with a logarithmic growth. The two curves are indistinguishable for the analyzed sizes and we cannot say which is the right scaling. The times grow for smaller connectivities but the scaling with $N$ remains the same.

\section{Replicated Belief Propagation}
\label{app:RBP}

\begin{figure}
    \centering
    \includegraphics[width=0.8\columnwidth]{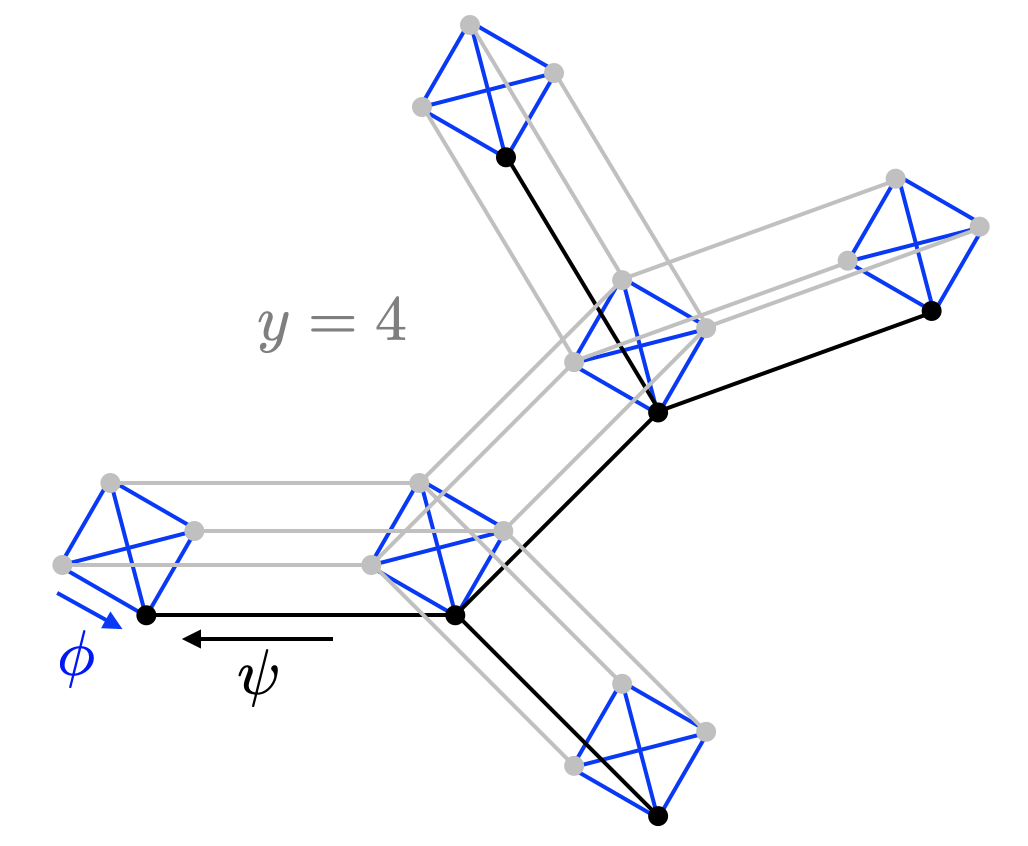}
    \caption{A schematic view of the replicated graph for $y=4$. The black graph is the original one, while the gray graphs are the $y-1$ replicas. Blue ferromagnetic couplings are present among every pair of replicated vertices with the same site index. Replicated BP uses two kind of messages, $\psi$ and $\phi$, represented by black and blue arrows.}
    \label{fig:RBP}
\end{figure}

In this appendix, we derive the Belief Propagation (BP) equations for the coloring problem with $y$ coupled replicas at inverse temperature $\beta$, which we call Replicated BP (RBP).
For this model, the probability distribution of the variables follows the Gibbs-Boltzmann probability associated with the Hamiltonian in Eq.~\eqref{eq:RSA_H}. As for the original model, one would like to estimate the marginal probabilities through the fixed point of some BP-like equations. However, in the replicated case, the graph is no more locally tree-like, because of the presence of small loops between the $y$ coupled replicas and the BP equations are not guaranteed to provide the correct solution to the problem.
We will see that in the limit $y\gg1$, the interaction among replicas is weak enough and the critical temperatures predicted by RBP will describe perfectly the transitions observed in the Replicated Simulated Annealing (RSA) in the limit of large $y$.

Making reference to Fig.~\ref{fig:RBP}, where every site of the replicated graph is identified by a pair $(i,a)$, with $i$ being the vertex of the original graph, taking values in $[1,N]$, and $a$ is the replica index running in $[1,y]$, the RBP equations for the coloring model with $y$ coupled replicas at inverse temperature $\beta$ can be written straightforwardly \footnote{In principle, when writing (linearized) BP equations, one should specify the iteration time indices $\tau$. However, having already many indices in Eqs.~(\ref{eq:repBP}), (\ref{eq:repBP_RS}), (\ref{eq:BPrep_lin}) and following, we omit the time indices, reminding the careful reader that time index is $\tau+1$ on the LHS and $\tau$ on the RHS.}
\begin{widetext}
\begin{equation} \label{eq:repBP}
\begin{split}
\psi^{(i,a)\to (j,a)}_s=&\frac{1}{Z^{(i,a)\to (j,a)}}\prod_{b(\neq a)} \left(1-(1-e^{\gamma\beta/y})\phi^{(i,b)\to (i,a)}_s\right)\prod_{k\in\partial i \backslash j} \left(1-(1-e^{-\beta})\psi^{(k,a)\to (i,a)}_s\right)\;,\\
\phi^{(i,a)\to (i,b)}_s=&\frac{1}{Z_{\phi}^{(i,a)\to (i,b)}}\prod_{c(\neq a,b)} \left(1-(1-e^{\gamma\beta/y})\phi^{(i,c)\to (i,a)}_s\right)\prod_{k\in\partial i} \left(1-(1-e^{-\beta})\psi^{(k,a)\to (i,a)}_s\right)\;,
\end{split}
\end{equation}
with $Z$ and $Z_{\phi}$ being the normalization factors enforcing $\sum_s \psi^{(i,a)\to (j,a)}_s = \sum_s \phi^{(i,a)\to (i,b)}_s=1$ for every edge of the replicated graph.
We call $\psi$ the cavity messages passed between nodes of the same replica, and $\phi$ the cavity messages passed between different replicas on the same vertex of the original graph. In the limit $y=1$, the $\phi$ messages are irrelevant and the above equations reduce to the standard BP equations for the original (unreplicated) model.

To simplify Eqs.~\eqref{eq:repBP} we assume a Replica Symmetric (RS) ansatz, where cavity messages do not depend on the replica indices, that is $\psi_s^{(i,a)\to(j,a)}=\psi_s^{i\to j}$ and $\phi_s^{(i,a)\to(i,b)}=\phi_s^i$, and then we get
\begin{equation} \label{eq:repBP_RS}
\begin{split}
\psi^{i\to j}_s=&\frac{1}{Z^{i\to j}} \left(1-(1-e^{\gamma\beta/y})\phi^i_s\right)^{y-1} \prod_{k\in\partial i \backslash j} \left(1-(1-e^{-\beta})\psi^{k\to i}_s\right)\;,\\
\phi^i_s=&\frac{1}{Z_{\phi}^i} \left(1-(1-e^{\gamma\beta/y})\phi^i_s\right)^{y-2}\prod_{k\in\partial i} \left(1-(1-e^{-\beta})\psi^{k\to i}_s\right)\;,
\end{split}
\end{equation}
\end{widetext}
We have verified numerically that, fixing $\gamma=1$, and solving Eqs.~\eqref{eq:repBP} and \eqref{eq:repBP_RS} on any given graph, we get exactly the same result, both in the high-temperature PM state and in the low-temperature FM state. So, the RS assumption is valid in the present case and we will focus only on Eqs.~\eqref{eq:repBP_RS} hereafter.

\section{Computation of the Kesten-Stigum transition temperature in the replicated model}
\label{app:KS}

The Kesten-Stigum transition temperature corresponds to the spinodal point of the PM state, that is to the lowest temperature such that the PM state is locally stable.
If the trivial fixed point of the RBP equations ($\psi_s=\phi_s=\frac{1}{q}$) is correctly describing the PM state of the replicated model, we can compute $T_{KS}(y,\gamma)$ via the study of the local stability of the RBP trivial fixed point with respect to fluctuations leading to the planted FM state. 

We perform a linear expansion around the PM trivial solution in the variables
\begin{equation}
    \epsilon_s^{i \to j}\equiv\psi_s^{i\to j}-\frac{1}{q}\;, \qquad 
    \delta_s^i\equiv\phi_s^{i}-\frac{1}{q}\;.
\end{equation}
At leading order in this expansion, Eqs.~\eqref{eq:repBP_RS} become
\begin{widetext}
\begin{equation} \label{eq:BPrep_lin}
\begin{split}
\epsilon_s^{i \to j}=&\sum_{k\in\partial i \backslash j}\sum_t \frac{\partial\psi_s^{i\to j}}{\partial\psi_t^{k \to i}}\bigg\rvert_\text{PM}\epsilon_t^{k\to i}+\sum_t \frac{\partial\psi_s^{i\to j}}{\partial\phi_t^{i}}\bigg\rvert_\text{PM}\delta_t^{i}=
A\sum_{k\in\partial i \backslash j}\left[\sum_{t}\epsilon_t^{k\to i }-q \epsilon_s^{k\to i }\right]+B(y-1)\left[\sum_t \delta^i_t -q \delta_s^i\right]\;,\\
\delta_s^{i}=&\sum_{k\in\partial i}\sum_t \frac{\partial\phi_s^{i}}{\partial\psi_t^{k \to i}}\bigg\rvert_\text{PM}\epsilon_t^{k\to i}+\sum_t \frac{\partial\phi_s^{i}}{\partial\phi_t^{i}}\bigg\rvert_\text{PM}\delta_t^{i}=A\sum_{k\in\partial i}\left[\sum_{t}\epsilon_t^{k\to i }-q \epsilon_s^{k\to i }\right]+B(y-2)\left[\sum_t \delta^i_t -q \delta_s^i\right]\;,
\end{split}
\end{equation}
\end{widetext}
where the subscript PM indicates that the function should be computed at the PM trivial solution $\psi_s=\phi_s=\frac{1}{q}$ and
\begin{equation*}
    A=\frac{1-e^{-\beta}}{q(q-1+e^{-\beta})}\;,\qquad B=\frac{1-e^{\gamma\beta/y}}{q(q-1+e^{\gamma\beta/y})}\;.
\end{equation*} 
Eqs.~\eqref{eq:BPrep_lin}) define a $q(N+M)\times q(N+M)$ matrix associated to a given graph of $N$ vertices and $M$ edges. To study the stability of the paramagnetic solution one should look at its maximum eigenvalue $\lambda_\text{max}$. 
When $\lambda_\text{max}=1$ the PM solution loses its local stability. In this way we can locate $\TKS(y,\gamma)$, that we have shown (fixing $\gamma=1$) in Fig.~\ref{Fig:phaseDiagram_y1} for $y=1$, in Fig.~\ref{Fig:New_diagram} for $y>1$ and in Fig.~\ref{Fig:RSAc17} with a vertical dashed line in the $y\gg 1$ limit. The latter value is a very good estimate of the jump temperature for large $y$, where the planted signal can be detected.

At this point, one could think to obtain an analytic estimation for $\TKS(y,\gamma)$ computing the eigenvalues of the linear transformation in Eqs.~(\ref{eq:BPrep_lin}), assuming that in the large $N$ limit the original graph becomes locally a tree, and so the cavity messages on the right-hand sides are uncorrelated.
This is a standard argument, that indeed works for $y=1$ and provides the analytical value $\TKS(y=1)=-1/\log\left[\frac{c-(q-1)^2}{q-1+c}\right]$ of Ref.~\cite{krzakala2009hiding}. 

In the following, we show that this kind of computation for $y>1$ does not give the correct result compared to the large $N$ limit estimation of $\TKS$ from the single graphs illustrated before.
Let us introduce $P_{s,t}(\epsilon^{i\to j})$ and $Q_{s,t}(\delta^{(i,a)\to (i,b)})$ as the probability distributions of the two kind of perturbations (between the nodes and between the replicas) in the direction of color $s$ if the spin $i$ was planted in the color $t$.
In the large $N$ limit, assuming a locally tree-like replicated graph, we can write closed equations for these distributions
\begin{widetext}
\begin{equation}
\begin{split}
P_{s,t}(\epsilon)=\sum_{d=0}^\infty \frac{d}{c}r_d \int \sum_{\{\sigma_j\},\{\tau_j\},\{\sigma_a\}}\prod_{j=1}^{d-1}d\epsilon_j P_{\sigma_j,\tau_j}(\epsilon_j)p(\tau_j|t)\prod_{a=1}^{y-1}d\delta_aQ_{\sigma_a,t}(\delta_a)\;
\delta(\epsilon-f_1(\{\epsilon_j\},\{\delta_a\}))\;,\\
Q_{s,t}(\delta)=\sum_{d=0}^\infty r_d \int \sum_{\{\sigma_j\},\{\tau_j\},\{\sigma_a\}}\prod_{j=1}^{d}d\epsilon_j P_{\sigma_j,\tau_j}(\epsilon_j)p(\tau_j|t)\prod_{a=1}^{y-2}d\delta_aQ_{\sigma_a,t}(\delta_a)\;\delta(\delta-f_2(\{\epsilon_j\},\{\delta_a\}))\;,
\end{split}
\end{equation}
\end{widetext}
where $r_d$ is the probability of having a vertex of degree $d$ in the original graph (a Poisson distribution of mean $c$ in the model we have studied in this work), $\frac{d}{c}r_d$ is the probability that, taken a link $i\to j$ at random, $i$ has degree $d$, $p(\tau|t)=\frac{\mathbb{I}(\tau\neq t)}{q-1}$ is the probability that a neighbour of site $i$ was planted in the state $\tau$ given that $i$ was planted in the state $t$, and finally the functions $f_1$ and $f_2$ are defined respectively by the right hand sides of the first and second equation in \eqref{eq:BPrep_lin}.
Please note that we are allowing $\delta_a$ to take different values for each replica. We are thus analyzing the linear expansion of Eq.~(\ref{eq:repBP}) rather than Eq.~\eqref{eq:BPrep_lin} which is the linear expansion of Eq.~(\ref{eq:repBP_RS}).

\begin{figure}[t]
\begin{center}
\includegraphics[width=\columnwidth]{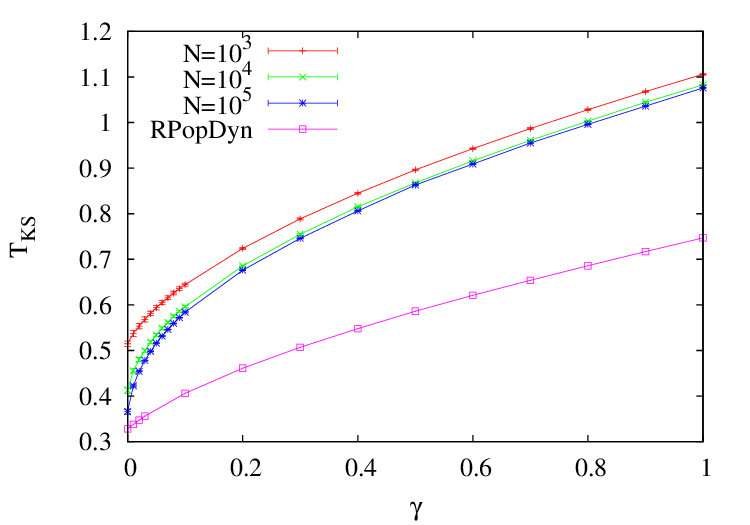}
\end{center}
\caption{$\TKS(y=64,\gamma)$ at $c=17$ as a function of the coupling $\gamma$ among replicas. The upper curves have been obtained running RBP on several graph of a given size $N$, while the lower curve has been obtained running population dynamics to solve Eqs.~\eqref{eq:BPrep_lin} or equivalently computing the largest eigenvalue of matrix in \eqref{eq:matrix}.
The limit $N\to\infty$ of the former computation coincides with the latter only at $\gamma=0$.
As soon as $\gamma>0$, the assumption of the replicated graph being locally tree-like is wrong and the latter computation fails, while running RBP on a given graph still provides a reliable value for $\TKS$.}
\label{Fig:RBP_vs_RPD}
\end{figure}

At this point, given that we are looking for a transition towards a "ferromagnetic" state, we can restrict our attention to the first moments of the above distributions
\begin{widetext}
\begin{equation}
\begin{split}
\<\epsilon_{s,t}\> & \equiv\int d\epsilon\,P_{s,t}(\epsilon)\, \epsilon = \sum_{d=0}^\infty \frac{d}{c}r_d \left\{\sum_{\{\tau_j\}\neq t}\frac{A}{q-1}\sum_{j=1}^{d-1}\left[\sum_{\sigma_j}\<\epsilon_{\sigma_j,\tau_j}\>-q\<\epsilon_{s,\tau_j}\>\right]+B\sum_{a=1}^{y-1}\left[\sum_{\sigma_a}\<\delta_{\sigma_a,t}\>-q\<\delta_{s,t}\>\right]\right\}\;,\\
\<\delta_{s,t}\> & \equiv\int d\delta\, Q_{s,t}(\delta)\, \delta = \sum_{d=0}^\infty r_d \left\{\sum_{\{\tau_j\}\neq t}\frac{A}{q-1}\sum_{j=1}^{d}\left[\sum_{\sigma_j}\<\epsilon_{\sigma_j,\tau_j}\>-q\<\epsilon_{s,\tau_j}\>\right]+B\sum_{a=1}^{y-2}\left[\sum_{\sigma_a}\<\delta_{\sigma_a,t}\>-q\<\delta_{s,t}\>\right]\right\}\;.
\end{split}
\end{equation}
\end{widetext}
Exploiting the symmetry under permutation of the colors, we can identify the only four relevant observables: $\epsilon^e\equiv \<\epsilon_{s,s}\>$, $\epsilon^d\equiv \<\epsilon_{s,t}\>$ with $s\neq t$, $\delta^e\equiv \<\delta_{s,s}\>$ and $\delta^d\equiv \<\delta_{s,t}\>$ with $s\neq t$. Our goal is to look at fluctuations towards the planted state, thus it is convenient to look at the differences $\epsilon^d-\epsilon^e$ and $\delta^d-\delta^e$, that satisfy the following closed equations
\begin{equation}
\begin{split}
\epsilon^d-\epsilon^e &= (\epsilon^d-\epsilon^e)A c \frac{q}{q-1} -q B (y-1) (\delta^d-\delta^e)\;,\\
\delta^d-\delta^e &= (\epsilon^d-\epsilon^e)A c \frac{q}{q-1} -q B (y-2) (\delta^d-\delta^e)\;.
\end{split}
\end{equation}
The analysis of the stability of the PM solution in the tree-limit of the replicated graph is thus reduced to the computation of the largest eigenvalues of the matrix
\begin{equation}
\begin{bmatrix}
    Ac\frac{q}{q-1} & -qB(y-1)\\
    Ac\frac{q}{q-1} & -qB(y-2)
\end{bmatrix}    
\label{eq:matrix}
\end{equation}
The value of $\TKS$ resulting from this computation for $y=64$ is shown in Fig.~\ref{Fig:RBP_vs_RPD} with the name `RPopDyn', as we have checked it coincides with the result found by running the population dynamic algorithm to solve Eqs.~(\ref{eq:BPrep_lin}). The comparison with the values of $\TKS$ obtained from the solution of the RBP equations on several graph realizations for different sizes $N$ reveals that in the large $N$ limit the two computations coincide only for $\gamma=0$, that is when replicas do not interact.
As long as $\gamma>0$ and the replicas are coupled, the large $N$ limit of the single graph computation does not coincide with the analytical estimation obtained under the assumption of a locally tree-like replicated graph.

The reason for the mismatch between the two computations is that for $y>1$ and $\gamma>0$ the replicated graph is not locally tree-like and cavity messages are correlated.
While running RBP on a given graph these correlations can be taken into account, the population dynamics algorithm, by construction, ignores any correlation among cavity messages. 
The correlations among messages coming from different replicas are positive, thanks to the ferromagnetic coupling among replicas, and this, in turn, produces a much larger value for $\TKS$ than under the assumption of uncorrelated messages.

The attentive reader could be puzzled by the fact that Population Dynamics gives the wrong result even in the $y\to\infty$ limit (the curves shown in Fig.~\ref{Fig:RBP_vs_RPD} are for $y=64$ that is quite large and their value is almost the asymptotic one for $y=\infty$). In fact in the limit $y\to\infty$ the $y$ replicas form a fully-connected ferromagnetic Potts model, and we know that the cavity method becomes exact for fully connected graphs. 
However, there is a simple explanation for the failure of Population Dynamics even when $y\to\infty$ in this case.
The Bethe approximation between the $y$ replicas is exact in the limit $y\to\infty$ \textit{inside} a single pure state. In fact the clustering property holds inside a pure state. Our model has a permutation symmetry: when the system is magnetized there are in principle different pure states, that can be obtained one from the other just by the permutation of the colors. Depending on which pure state is chosen by the $y$-replicas on the site $i$, also the neighboring spins $j\in \partial i$ will choose their color. In the replicated graph there are many small loops of length 4, between the spins $(i,a)-(i,b)-(j,b)-(j,a)$. While in the limit $y\to\infty$ the connected correlation functions between $(i,a)$ and $(i,b)$ and between $(j,a)$ and $(j,b)$ decay to zero, the full correlation functions do not decay to zero, because the average magnetization is not zero inside a pure state. The BP algorithm for single instance of the graph takes into account this correlation of the magnetizations, just because, when applied to a fixed instance of the graph, BP spontaneously breaks the permutation of the colors just dynamically choosing a given pure state, and thus it is exact in the limit of $y\to\infty$. On the other hand, the Population Dynamics algorithm (Eqs.~C3) does not consider the correlation in the magnetizations, picking each time different neighboring spins $i$ and $j$ as if they were uncorrelated, and this random choice is wrong \textit{even} in the limit $y\to\infty$.

Willing to improve the analytical computation (or the solution via the population dynamics algorithm) one should consider a \emph{super-variable} at every node of the original graph, taking $q^y$ values. In this case, the graph connecting the super-variables is indeed locally tree-like and a population dynamical computation would give the correct estimates for the thermodynamical thresholds both for finite and infinite $y$ (we leave this computation for future work).

\section{Computation of the dynamical transition temperature}
\label{app:Td}

The dynamical temperature $T_d$ is the temperature below which glassy states are locally stable (and also attractive for the out of equilibrium dynamics). $T_d$ is not modified by the addition of the planted state, which is roughly orthogonal to the glassy states. For this reason, the usual procedure to identify $T_d$ is as follows: one plants a configuration typical of the Gibbs measure at temperature $T$ constructing a graph compatible with it. This is a generalization of the planting procedure at $T=0$ that we have described in the main text.
In practice, to plant a configuration of temperature $T$ when $y=1$ one first decides the assigned color $s_i$ for each node $i$, in such a way that there are $N/q$ nodes for each color. Then one extracts two nodes $i$ and $j$ at random among the $N$ possible ones, and an edge is put between them with 
probability $p=e^{-\beta (1-\delta_{s_i,s_j})}$. In practice non-monocromatic edges are always accepted, while monochromatic edges are accepted with probability $e^{-\beta}$ that tends to zero if $T\to 0$. One then repeats this operation until $M=\frac{N c}{2}$ edges are put in the graph.
The chosen planted configuration is in this way an equilibrium configuration of the constructed graph at temperature $T$.
One then initializes BP near enough to the planted glassy configuration, and runs the BP equations at temperature $T$. If the fixed point reached by BP is the PM one then glassy states are locally unstable and thus $T>T_d$, while if the fixed point is correlated with the planted glassy state, then $T<T_d$.

In this way we have located $T_d$ for $y=1$ (see data in Fig.~\ref{Fig:phaseDiagram_y1}).
\footnote{One could locate $T_d$ also performing a standard 1RSB computation, looking when a 1RSB solution is stable and different from the PM one, however, this procedure is much more expensive than the planting procedure we used.}

\begin{figure}[t]
\begin{center}
\includegraphics[width=\columnwidth]{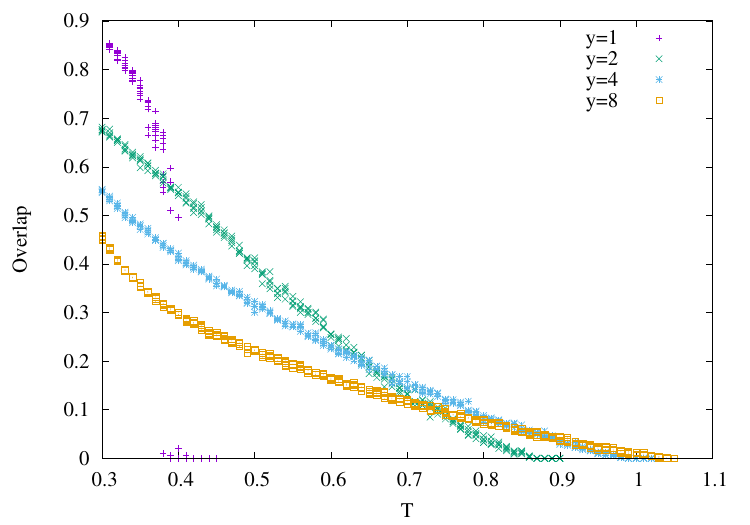}
\end{center}
\caption{Overlap between the fixed point reached by BP initialized near a planted glassy state at $T$ and the glassy configuration planted at $T$, as a function of
the planting temperature $T$ for different values of the replicas $y$ at $c=17$, $N=10^4$, $\gamma=1$. At each temperature the results for 5 different planted graphs are shown. $T_d$ is located as the lowest temperature at which the overlap reaches 0.
As long as $y>1$, the transition becomes continuous.}
\label{Fig:overlap}
\end{figure}

\begin{figure}[t]
\begin{center}
\includegraphics[width=\columnwidth]{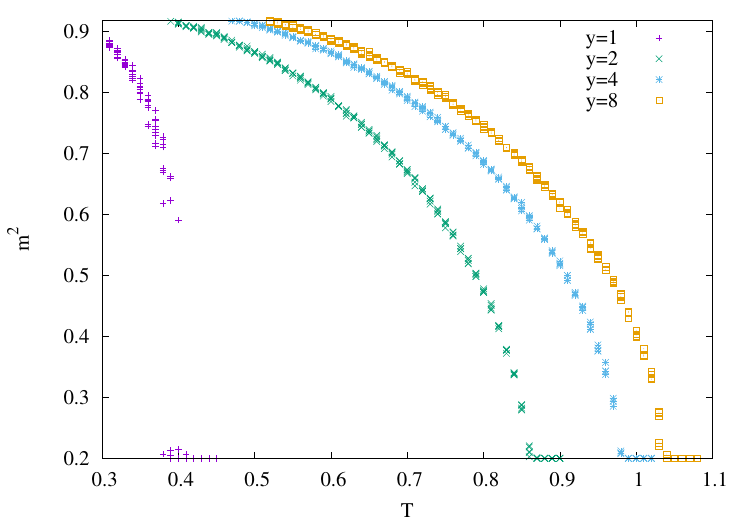}
\end{center}
\caption{Squared magnetization defined as in eq. (\ref{eq:m2}) of the fixed point reached by BP initialized near to a planted glassy state at $T$, as a function of
the planting temperature $T$ for different values of the replicas $y$ at $c=17$, $N=10^4$, $\gamma=1$. At each temperature the results for 5 different planted graphs are shown. $T_d$ is located as the lowest temperature at which $m^2=\frac{1}{q}$.
As long as $y>1$, the transition becomes continuous.}
\label{Fig:m2}
\end{figure}

In principle one could use the same planting technique to locate $T_d$ in the $y>1$ case, paying attention to plant a state according to the replicated Hamiltonian in Eq.~\eqref{eq:RSA_H}. In practice we did the following: we first decided the assigned color $s_i$ for each node $i$, in such a way that there are $N/q$ nodes for each color. We put all the $y$ replicas associated to the node $i$ in the color $s_i$ at the beginning. Then we run a short Monte Carlo at temperature $T$ for the $y$ replicas of node $i$ with associated Hamiltonian $H_i=-\frac{\gamma}{2y}\sum_{a\neq b=1}^y \delta_{s_i^a, s_i^b}$. In this way we obtain an equilibrium configuration for the $y$ replicas of each of the $N$ nodes (that at this point are still disconnected). Then we construct the graph: we extract two nodes $i$ and $j$ at random among the $N$ possible ones, and an edge is put between them with probability $p=e^{-\beta \sum_{a=1}^y(1-\delta_{s_i^a,s_j^a})}$. We then repeat this operation until $M=\frac{N c}{2}$ edges are put in the graph.
To test the local stability of this planted configuration, one really needs to run the BP algorithm with $y$ different replicas, according to Eqs.~\eqref{eq:repBP}, because the different replicas will not be planted necessarily with the same color (being the temperature positive $T>0$ each replica will have a non zero probability to differ from the planted configuration). Computing $T_d$ in this way, we have discovered that the transition becomes a continuous one as soon as $y>1$ (at least for $\gamma=1$). This is signaled by the fact that the overlap between the fixed point reached by BP initialized near the planted glassy state at $T$ and the planted glassy configuration tends to zero continuously in the limit $T\to T_d^-$. 
One can also look at another order parameter defined as
\begin{equation}
m^2=\frac{1}{yN}\sum_{i=1}^N \sum_{a=1}^y\sum_{s=1}^q (\psi^{(i,a)}_s)^2.
\label{eq:m2}
\end{equation}
In the paramagnetic state, $\psi^{(i,a)}_s=\frac{1}{q}\quad \forall s,i,a$, and thus $m^2=\frac{1}{q}$. When $y=1$, $m^2$ jumps discontinuously to the value $m^2=\frac{1}{q}$ at $T=T_d$, while, for $y>1$, $m^2$ reaches its paramagnetic value continuously at $T_d$.

Both the behaviour of the overlap and the squared magnetization, summarized in Figs.~\ref{Fig:overlap} and \ref{Fig:m2}, imply that for $y>1$ the dynamical transition of the random model changes nature and becomes continuous, at difference with the case $y=1$, where this transition is discontinuous.
This fact implies that we can compute $T_d$ in the $y>1$ case just by looking at the instability of the paramagnetic solution through the computation of the maximal eigenvalue of Eqs.~(\ref{eq:BPrep_lin}) on a given graph \emph{without} the $T=0$ planted solution  (exactly as we did to compute $\TKS$ on a graph with the $T=0$ planted solution). We checked that this computation gives exactly the same result for $T_d$ as the previously described planting procedure. However the study of the linearized equations is much faster, the convergence of BP initialized around the planted glassy state being very slow in the vicinity of $T_d$. 
With this linearization method we have computed the $T_d$ values for $y>1$ shown in Fig.~\ref{Fig:New_diagram}.

\bibliography{biblio}

\end{document}